\documentclass[aps,prx,twocolumn,amsmath,amssymb,superscriptaddress,floatfix,reprint,longbibliography]{revtex4-1}
\usepackage{epsfig}
\usepackage{amsmath}
\usepackage{amssymb}
\usepackage{amsfonts}
\usepackage{mathptmx}
\usepackage{dcolumn}
\usepackage{eucal}
\usepackage{bm}
\usepackage{color}
\usepackage[colorlinks,linkcolor=blue,citecolor=blue]{hyperref}
\usepackage{epstopdf}
\usepackage{url}
\usepackage{svg}
\usepackage{braket}

\begin{document}
\title{Orbital design of Berry curvature: pinch points and giant dipoles induced by crystal fields}

\author{Maria Teresa Mercaldo}
\affiliation{Dipartimento di Fisica ``E. R. Caianiello", Universit\`a di Salerno, IT-84084 Fisciano (SA), Italy}

\author{Canio Noce}
\affiliation{Dipartimento di Fisica ``E. R. Caianiello", Universit\`a di Salerno, IT-84084 Fisciano (SA), Italy}
\affiliation{SPIN-CNR, IT-84084 Fisciano (SA), Italy}

\author{Andrea D. Caviglia}
\affiliation{Department of Quantum Matter Physics, University of Geneva, 24 Quai Ernest Ansermet, CH-1211 Geneva, Switzerland}

\author{Mario Cuoco}
\affiliation{SPIN-CNR, IT-84084 Fisciano (SA), Italy}
\affiliation{Dipartimento di Fisica ``E. R. Caianiello", Universit\`a di Salerno, IT-84084 Fisciano (SA), Italy}

\author{Carmine Ortix}
\affiliation{Dipartimento di Fisica ``E. R. Caianiello", Universit\`a di Salerno, IT-84084 Fisciano (SA), Italy}

\begin{abstract}
\noindent
The Berry curvature (BC) -- a quantity encoding the
geometric properties of the electronic wavefunctions in a solid -- is at the heart of 
different Hall-like transport phenomena, including 
the anomalous Hall 
and the non-linear Hall and Nernst effects. 
In non-magnetic quantum materials with acentric crystalline arrangements, local concentrations of BC are generally linked to 
single-particle wavefunctions that 
are a quantum superposition of electron and hole excitations. 
BC-mediated effects are consequently observed in two-dimensional systems
with pairs of massive Dirac cones 
and 
three-dimensional bulk crystals with
quartets of Weyl cones.
Here, we demonstrate that 
in materials equipped with orbital degrees of freedom
local BC concentrations can arise even 
in the complete absence of hole excitations. 
In these solids, the crystals fields 
appearing in very low-symmetric structures
trigger BCs characterized by hot-spots and singular pinch points.
These characteristics naturally yield giant BC dipoles and large non-linear transport responses in time-reversal symmetric conditions. 
\end{abstract}
\maketitle

\section{Introduction} 
Quantum materials can be generally defined as those solid-state structures hosting physical phenomena which, even at the macroscopic scale, cannot be captured by a purely classical description~\cite{kei17}.
Among such quantum phenomena, those related to the geometric properties of the electronic wavefunctions play undoubtedly a primary role. 
In an $N$-band crystalline system, 
the cell-periodic part of the electronic Bloch waves 
defines a mapping from the Brillouin zone (BZ) to 
a complex space
naturally equipped with a geometric structure --  its tangent space defines a Fubini-Study metric~\cite{ana90} that measures the infinitesimal distance between Bloch states at different points of the BZ. The imaginary part of this quantum geometric tensor~\cite{pro80,kah21} corresponds to the well-known Berry curvature (BC), which, when integrated over the full BZ, gives the Chern number cataloguing  two-dimensional insulators~\cite{Thouless}. In metallic systems with partially filled bands, the BC summed over all occupied states can result in a non-vanishing Berry phase if the system breaks time-reversal symmetry. This Berry phase regulates the intrinsic part of the anomalous Hall conductivity of magnetic metals~\cite{Haldane2,nag10,Gro20,vanthiel21}.

Materials with an acentric crystal structure can possess non-vanishing concentrations of 
BC even if magnetic order is absent. Probing the BC of these non-centrosymmetric and non-magnetic materials via charge transport measurements usually requires externally applied magnetic fields. For instance, in time-reversal invariant Weyl semimetals, such as TaAs~\cite{lv15}, the strong 
BC arising from the Weyl nodes 
can be revealed using the planar Hall effect~\cite{nan17} -- a physical  consequence of the negative longitudinal magnetoresistance associated with the chiral anomaly of Weyl fermions~\cite{arm18}. Recently, it has been also shown that the planar Hall effect can display an anomalous antisymmetric response~\cite{bat21,cul21}, which, at least in two-dimensional materials, is entirely due to an unbalance in the BC distribution triggered by the Zeeman-induced spin splitting of the electronic bands. 

In the absence of external magnetic fields, a BC charge transport diagnostic for non-magnetic materials requires to go beyond the linear response regime~\cite{dey09,moo10,Sodemann2015,Ortix2021,du19}. Hall-like currents appearing as a non-linear (quadratic) response to a driving electric field can have an intrinsic contribution governed by the Berry curvature dipole (BCD), which is essentially the first moment of the Berry curvature in momentum space. 
In three-dimensional systems, non-vanishing BCDs have been linked to the presence of tilted Weyl cones, and have been shown to exist both in type-I 
and in type-II Weyl semimetals~\cite{sol15} such as MoTe$_2$~\cite{sin20} and the ternary compound TaIrTe$_4$~\cite{koe16,kum21}. Furthermore, the Rashba semicondutor BiTeI has been predicted to host a BCD that is strongly enhanced across its pressure-induced topological phase transitions~\cite{fac18}. 

In two-dimensional materials, the appearance of BCDs is subject to stringent symmetry constraints: the largest symmetry group is ${\mathcal C}_s$, which is composed by the identity and a single vertical mirror line \footnote{Note that non-linear Hall currents can exist in symmetry groups containing also rotational symmetries. In these cases non-linear skew and side-jump scatterings are the origin of the phenomenon}. 
The concomitant 
presence of spin-orbit coupled massive Dirac cones with substantial BC and such unusually low-symmetry crystalline environments have suggested the surface states of SnTe~\cite{TCI} in the low-temperature ferroelectric phase~\cite{lau19}, monolayer transition metal dichalcogenides in the so-called $1T_{d}$ phase~\cite{xu18,You2018,kan19}, and bilayer WTe$_2$ as material structures hosting sizable BCDs~\cite{ma19, du18}.  
Spin-orbit free two-dimensional materials, including monolayer and bilayer graphene, have been also put forward as materials with relatively large BCDs~\cite{ho21}. 
In these systems, it is the 
interplay
between the trigonal warping of the Fermi surface and the presence of massive Dirac cones due to inversion symmetry breaking 
that triggers dipolar concentration of Berry curvatures~\cite{bat19}. 

Finite concentrations of BC and BCDs are symmetry allowed also in systems that do not feature quartets of Weyl cones and pairs of massive Dirac cones. 
The anomalous massless Dirac cones at the surface of three-dimensional strong topological insulators~\cite{he21} as well as
conventional two-dimensional electron gases (2DEG) with Rashba spin-orbit coupling~\cite{les22} are generally characterized by finite local BC concentrations when subject to trigonal crystal fields. The existence of 
BC in 2DEGs,
which has been experimentally probed through 
``anomalous" planar Hall effect measurements~\cite{les22}, provides a new avenue for investigations. 
It shows in fact that Berry curvature-mediated effects can be generated entirely from conduction electrons. This overcomes the requirement of materials with narrow gaps in which the electronic wavefunctions at the Fermi level 
are a quantum superposition of electron and hole excitations,
and extends the palette of non-magnetic materials displaying BC effects to, for instance, doped semiconductors with gaps in the eV range.
It also proves that it is possible to trigger BC effects in conventional electron liquids with competing instabilities towards other many-body quantum phases. 

In a spin-orbit coupled 2DEG, the BC is however triggered by crystalline anisotropy terms, which are cubic in momentum and linked to the out-of-plane component of the spin textures~\cite{he18,tra22}. 
Consequently, the BC does not possess the characteristic ``hot-spots" appearing in close proximity to near degeneracy between two bands where the Bloch wavefunctions are rapidly changing. The absence of such BC hot-spots forbids, in turn, large enhancements of the BCD, which is a central quest for material design.  
This motivates the fundamental question on whether and how an electron system can develop strong local BC concentrations in time-reversal symmetric conditions even in the complete asbence of hole excitations. 
Here, we provide a positive answer to this question by showing that 
spin-orbit free
metallic systems with an effective pseudo-spin one orbital degree of freedom can display BC hot-spots and characteristic BC singular pinch points  that yield dipoles order of magnitudes larger than those triggered by spin-orbit coupling in a 2DEG.

\begin{figure*}[tbp]
    \includegraphics[width=\textwidth]{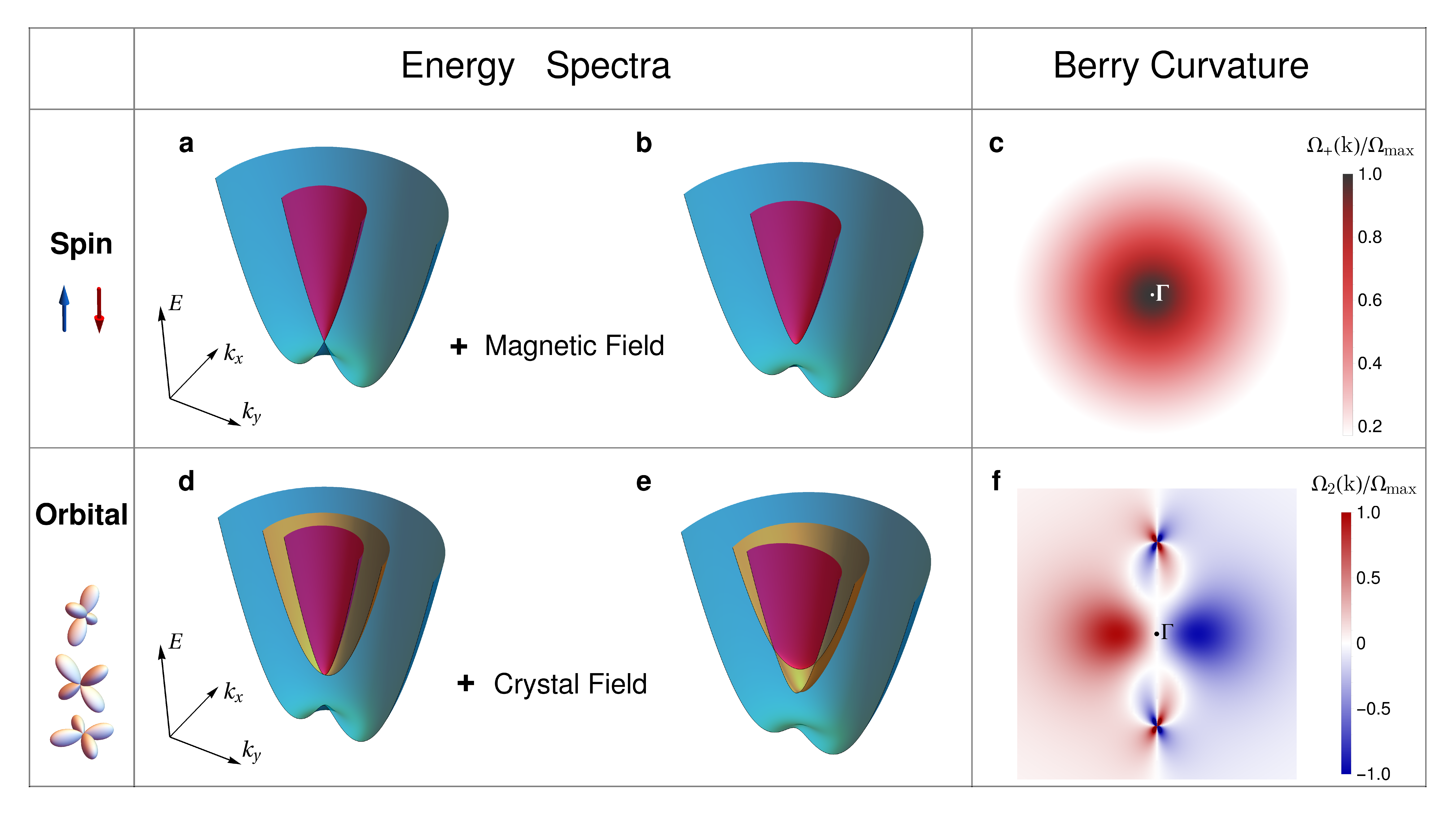}
    \caption{(a) Schematic band structure of a two-dimensional electron gas with Rashba spin-orbit coupling. (b) An out-of-plane magnetic field breaks the Kramers' degeneracy at ${\bf k}=0$ and triggers a finite BC. (c) The local BC has a circular profile with an hot spot at the $\Gamma$ point of the BZ. (d) Schematic band structure of a two-dimensional electron system characterized by an $L=1$ orbital multiplet in a trigonal crystalline environment. (e) An additional crystalline symmetry lowering splits completely the energy levels at the $\Gamma$ point of the BZ even if time-reversal symmetry is preserved. The presence of mirror symmetry protects crossing at finite momenta. (f) A characteristic time-reversal symmetric BC profile with the presence of hot-spots and singular pinch points. The BC has been obtained using the model Hamiltonian Eq.~\eqref{eq:Ham} with $\Delta=-0.2 {\cal E}_0$, $\Delta_m=0.12 {\cal E}_0$, $\alpha_R=1.0 {\cal E}_0/k_F^0$ and $\alpha_m=0.5 {\cal E}_0 / k_F^0$ with ${\cal E}_0=\hbar^2 (k_F^0)^2 / (2 m)$ and $k_F^0$ a characteristic Fermi wavevector.}
    \label{fig:fig1}
\end{figure*}

\section{Results}
\subsection{Model Hamiltonian from symmetry principles}
Let us first consider a generic single-valley two-level system in two dimensions with spin degree of freedom only. The corresponding energy spectrum is assumed to accurately represent the electronic bands close to the Fermi level of the metal in question. As long as we consider materials without long-range magnetic order, the two Fermi surfaces must originate from one of the four time-reversal invariant point of the Brillouin zone (BZ) $(n_1 {\bf b}_1 + n_2 {\bf b}_2)/2$ with ${\bf b}_{1,2}$ the two primitive reciprocal lattice vectors of the BZ and $n_{1,2}=0,1$ \footnote{We emphasize that our arguments hold also for systems with multiple pairs of pockets centered at time-reversal invariant momenta}. Time-reversal symmetry  guarantees that the two bands will be Kramers' degenerate at the time-reversal invariant momenta (TRIM). The effective Hamiltonian in the vicinity of the TRIM can be captured using a conventional ${\bf k} \cdot {\bf p}$ theory that keeps track of the point group symmetries of the crystal. To make things concrete, let us assume that the low-energy conduction bands are centered around the $\Gamma$ point of the BZ and we are dealing with an acentric crystal with ${\mathcal C}_{3v}$ point group symmetry. This is the largest acentric symmetry group without ${\mathcal C}_2 {\mathcal T}$ symmetry, ${\mathcal C}_2$ indicating a twofold rotation symmetry with out-of-plane axis and ${\mathcal T}$ time-reversal, and thus allows for local BC concentrations~\cite{bat21}. The generators of ${\mathcal C}_{3v}$ are the threefold rotation symmetry ${\mathcal C}_3$ and a vertical mirror symmetry, which, without loss of generality, we take as ${\mathcal M}_x$ sending $x \rightarrow -x$. The threefold rotation symmetry can be represented as $\mathrm{e}^{-i \pi \sigma_z / 3}$ while the mirror symmetry as $i \sigma_x$~\cite{fu09}. 
Momentum and spin transform under ${\mathcal C}_3$ and ${\mathcal M}_x$ as follows
\begin{align}
{\mathcal C}_3: \hspace{.1cm}  k_{\pm} \rightarrow \mathrm{e}^{\pm 2 \pi i /3} k_{\pm} ; \hspace{.2cm}& \sigma_{\pm} \rightarrow \mathrm{e}^{\pm 2 \pi i /3} \sigma_{\pm} & \sigma_z \rightarrow \sigma_z \nonumber \\
{\mathcal M}_x: \hspace{.1cm} k_{\pm} \rightarrow -k_{\pm} ; \hspace{.2cm} & \sigma_{y,z} \rightarrow -\sigma_{y,z} & \sigma_x \rightarrow \sigma_x 
\end{align}
where $k_{\pm}=k_x \pm i k_y$ and $\sigma_{\pm}=\sigma_x \pm i \sigma_y$. Furthermore, the Hamiltonian must satisfy the time-reversal symmetry constraint ${\mathcal H}({\bf k})={\mathcal T} {\mathcal H}(-{\bf k}) {\mathcal T}^{-1}$, with the time-reversal operator that, as usual, can be represented as ${\mathcal T}=i \sigma_y {\mathcal K}$ and ${\mathcal K}$ the complex conjugation. When expanded up to linear order in ${\bf k}$, the form of the Hamiltonian reads as ${\mathcal H}({\bf k})= \alpha_R \left(k_x \sigma_y - k_y \sigma_x \right)$. The Dirac cone energy spectrum predicted by this Hamiltonian violates the fermion doubling theorem~\cite{nie81} and hence can occur only on the isolated surfaces of three-dimensional strong topological insulators~\cite{has10}. And indeed  ${\mathcal H}({\bf k})$ coincides with the effective Hamiltonian for the surface states of the topological insulators in the Bi$_2$Se$_3$ material class~\cite{Zhang2009,3dTIexp,fu09}. In a genuine two-dimensional system  such anomalous states cannot be present, and an even number of Kramers' related pair of bands must exist at each Fermi energy. Consequently, the effective Hamiltonian must be equipped with an additional term that is quadratic in momentum and such that it doubles the number of states at each energy. Time-reversal symmetry implies that terms quadratic in momentum are coupled to the identity matrix. Therefore, we arrive at the well-known Hamiltonian of a two-dimensional electron gas with Rashba-like spin-orbit coupling that reads 
\begin{equation}
{\mathcal H}({\bf k})=\dfrac{\hbar^2 k^2}{2 m} \sigma_0 + \alpha_R \left(k_x \sigma_y - k_y \sigma_x \right). 
\label{eq:hamrashba2deg}
\end{equation}
The corresponding energy spectrum consisting of two shifted parabolas is schematically shown in Fig.~\ref{fig:fig1}(a). 
Although the crystalline symmetry requirements are fulfilled, the Hamiltonian above does not predict any finite BC local concentration. This is because the ${\bf d}$ vector associated to the Hamiltonian ${\bf d}=\left\{-\alpha_R k_y, \alpha_R k_x,0\right\}$ is confined to a two-dimensional plane at all momenta.

There are two different ways to lift the ${\bf d}$ vector out-of-plane and thus trigger a non-vanishing BC~\cite{xia10}. The first one consists in introducing a constant mass $\Delta \sigma_z$. This term removes the Kramers' degeneracy at the TRIM [see Fig.~\ref{fig:fig1}(b)] and therefore breaks time-reversal invariance. It can be realized by externally applying an out-of-plane magnetic field or by inducing long-range magnetic order. 
The BC then generally displays an hot-spot located at the TRIM and a circular symmetric distribution [see Fig.~\ref{fig:fig1}(c)].
Moreover, time-reversal symmetry breaking implies that 
the Berry phase accumulated by electrons on the Fermi surface is non-vanishing~\cite{Haldane2}.  The second route explicitly takes into account trigonal warping terms which are cubic in momentum and couple to the Pauli matrix $\sigma_z$. Such terms preserve time-reversal invariance, and thus create a BC distribution with an angular dependence such that the Berry phase accumulated over any symmetry-allowed Fermi line
cancels out~\cite{he21}. Perhaps more importantly,
the BC triggered by crystalline anisotropy terms~\cite{les22} does not display a hot-spot, thus suggesting that in systems with conventional quasiparticles and a single internal degree of freedom time-reversal symmetry breaking is a prerequisite for large local BC enhancements.

We now 
refute
this assertion by showing that in systems with orbital degrees of freedom the formation of BC hot-spots is entirely allowed even in time-reversal symmetric conditions. 
Consider for instance a system of $p$ orbitals. In a generic centrosymmetric crystal, interorbital hybridization away from the TRIM can only occur with terms that are quadratic in momentum. However, and this is key, in an acentric crystal interorbital mixing terms linear in momentum are symmetry allowed. These mixing terms, often referred to as orbital Rashba coupling \cite{Park11,Kim13,Mer20}, are able to induce BC hot spots with time-reversal symmetry, as we now show. 
We assume as before an acentric crystal with ${\mathcal C}_{3v}$ point group, 
and electrons that are effectively spinless due to $SU(2)$ spin symmetry conservation: we are thus removing spin-orbit coupling all together. 
In the $p_z,p_y,p_x$ orbital basis, the generators of the point group are represented by
\begin{equation*}
{\mathcal M}_x=\left( \begin{array}{ccc} 1 & 0 & 0 \\ 0 & 1 & 0 \\ 0 & 0 & -1 \end{array} \right) ; \hspace{.3cm} {\mathcal C}_3 = \left( \begin{array}{ccc} 1 & 0 & 0 \\  0 & \cos{\frac{2\pi}{3}} & \sin{\frac{2 \pi}{3}} \\ 0 & -\sin{\frac{2 \pi}{3}} & \cos{\frac{2 \pi}{3}} \end{array}
\right),
\end{equation*}
The two $p_{x,y}$ orbitals form a two-dimensional irreducible representation (IRREP) whereas the $p_z$ orbital represents a one-dimensional IRREP. 
The form of the effective Hamiltonian away from the TRIM can be captured using symmetry 
constraints. 
Specifically, any generic $3 \times 3$ Hamiltonian can be expanded in terms of the nine Gell-Mann matrices~\cite{bar12} $\Lambda_i$ [see Methods] as 
\begin{equation}
{\mathcal H}({\bf k})=\sum_{i=0}^8  b_i({\bf k}) \Lambda_i.
\end{equation}
The invariance of the Hamiltonian requires that the components of the Hamiltonian vector ${\bf b}({\bf k})$ should have the same behavior as the corresponding Gell-Mann matrices $\Lambda_i$. This means that they should belong to the same representation of the crystal point group~\cite{liu10}. From the representation of the $\Lambda_i$'s [see Methods and Table~1] and those of the polynomials of ${\bf k}$ [see Table~1], we find that the effective Hamiltonian up to linear order in momentum reads as
\begin{equation}
\label{eq:H0}
{\mathcal H}({\bf k})=\Delta \left(\Lambda_3 + \dfrac{1}{\sqrt{3}} \Lambda_8 \right) - \alpha_R \left[k_x \Lambda_5 + k_y \Lambda_2 \right].
\end{equation}
Here the parameter $\Delta$ quantifies the energetic splitting between the $p_{x,y}$ doublet and the $p_z$ singlet. 
The second term in the Hamiltonian corresponds instead to the pseudo-spin one massless Dirac Hamiltonian~\cite{gre10,gio15} predicted to occur for instance in the kagome lattice with a staggered magnetic $\pi$ flux~\cite{gre10}. 
Pseudo-spin one Dirac fermions are not subject to any fermion multiplication theorem~\cite{fan19}. Therefore, a doubling of the number of states at each energy is not strictly required. However, since we are interested in systems without the concomitant presence of electrons and holes, we will introduce a term $\hbar^2 k^2 \Lambda_0 / (2 m)$ with 
an equal effective mass
for all three  bands. The ensuing Hamiltonian can be then seen as a generalization of the Rashba 2DEG to an $SU(3)$ system with the effect of the trigonal crystal field that leads to a partial splitting of the energy levels at the TRIM, entirely allowed by the absence of Kramers' theorem. Despite the spectral properties [c.f. Fig.~\ref{fig:fig1}(d)] have a strong resemblance to those obtained in a time-reversal broken 2DEG, a direct computation [see Methods] shows that the BC associated to the Hamiltonian above is vanishing for all momenta. 
Breaking time-reversal symmetry introducing a constant mass term $\propto \Lambda_7 $ or considering crystalline anisotropy terms that are cubic in momentum represent two possible routes to trigger a finite Berry curvature. 
The crux of the story is that in the present $SU(3)$ system at hand, another possibility exists. 
It only relies on the crystal field effects that are generated by lowering the crystalline point group to ${\mathcal C}_s$. From the representations of the Gell-Mann matrices and the polynomials of ${\bf k}$ in this group, we find that the effective Hamiltonian 
reads
\begin{eqnarray}
\label{eq:Ham}
{\mathcal H}({\bf k})&=&\dfrac{\hbar^2 k^2}{2 m} \Lambda_0 + \Delta \left(\Lambda_3 + \dfrac{1}{\sqrt{3}}  \Lambda_8 \right) + \Delta_m \left(\dfrac{1}{2} \Lambda_3 -\dfrac{\sqrt{3}}{2} \Lambda_8 \right) \nonumber \\ & & - \alpha_R \left[k_x \Lambda_5 + k_y \Lambda_2 \right] - \alpha_m k_x \Lambda_7 .
\end{eqnarray}
Nothing prevents to have the interorbital mixing terms $\propto \Lambda_{2,5}$ with different amplitudes. Without loss of generality, in the remainder we will consider a single parameter $\alpha_R$. 
In the Hamiltonian above, we have also neglected a constant term $\propto \Lambda_1$. 
For materials with an high-temperature trigonal structure, its amplitude $\Delta_1$ is expected to be of the same order of magnitude as $\Delta_m$. In this regime [see the Supplemental Material], a term $\propto \Lambda_1$ has a very weak effect on the energy spectrum and BC properties, and can be thus disregarded~\footnote{At the center of the BZ, a constant term $\propto \Lambda_1$ mixes the $p_{y,z}$ states and consequently lifts the energy degeneracy of the $p_{x,y}$ doublet precisely as the constant term parametrized by $\Delta_m$ does. However, the splitting due to $\Lambda_1$ is strongly suppressed (see the Supplemental Material) by the energy difference between the $p_{y,z}$ states caused by the trigonal crystal field $\Delta$.}.
The energy spectrum reported in Fig.~\ref{fig:fig1}(e) shows that the effect of the crystal symmetry lowering is twofold. First, there is an additional energy splitting between the $p_{x,y}$ implying that all levels at the $\Gamma$ point of the BZ
are singly degenerate. Second, the two $p_{x,y}$ orbitals have band degeneracies along the mirror symmetric $k_x=0$ line of the 
BZ.
Such mirror-symmetry protected crossings give rise to BC singular pinch points [see Fig.~\ref{fig:fig1}(f) and the Supplemental Material]. 
It is the presence of these pinch points that represents
the hallmark of the non-trivial geometry of the electronic wavefunctions associated to the $p$-orbital manifold. Note that the BC also displays hot-spots [see Fig.~\ref{fig:fig1}(f)] with BC sources and sinks averaging to zero on any mirror symmetric Fermi surface as mandated by time-reversal invariance. 

\begin{table}
\begin{tabular}{|c|c|c|c|c|c|}
\hline 
${\mathcal C}_{3v}$ & E & 2 ${\mathcal C}_3$ & $2 \sigma_v$ & polynomials of ${\bf k}$ & Gell-Mann matrices 
\tabularnewline
\hline 
$A_1$ & $1$ & $1$ & $1$ & $1$, $k_x^2+k_y^2$ & $\Lambda_3 + \Lambda_8 / \sqrt{3}, \Lambda_0$ 
\tabularnewline 
\hline 
$A_2$ & $1$ & $1$ & $-1$ & -- & $\Lambda_7$
\tabularnewline  
\hline 
$E$ & $2$ & $-1$ & $0$ & $\left\{k_x, k_y \right\}$ &$\left\{\Lambda_1, \Lambda_4\right\}$, $\left\{\Lambda_2, \Lambda_5\right\}$ 
\tabularnewline 
& & & & & $\left\{\Lambda_6, \Lambda_3 / 2 - \sqrt{3} \Lambda_8 / 2  \right\}$ 
\tabularnewline 
\hline 
\end{tabular}

\vspace{1cm}
\begin{tabular}{|c|c|c|c|c|}
\hline 
${\mathcal C}_{s}$ & E & $2 \sigma_v$ & polynomials of ${\bf k}$ & Gell-Mann matrices 
\tabularnewline
\hline 
$A^{\prime}$ & $1$ & $1$ & $1$, $k_y$, $k_x^2$, $k_y^2$ & $\Lambda_1$, $\Lambda_2$, $\Lambda_3$, $\Lambda_8$ 
\tabularnewline 
\hline 
$A^{\prime \prime}$ & $1$ & $-1$ & $k_x$ & $\Lambda_4$, $\Lambda_5$, $\Lambda_6$, $\Lambda_7$ 
\tabularnewline  
\hline 
\end{tabular}
\caption{Character table for the point groups ${\mathcal C}_{3v}$ and ${\mathcal C}_s$. We also indicate the representation of the Gell-Mann matrices and the polynomials of momentum ${\bf k}$. The model Hamiltonians reported in the main text can be obtained by additionally using the time-reversal symmetry constraint ${\mathcal H}^{\star}(-k_x,-k_y)={\mathcal H}(k_x,k_y)$.}
\label{tab:tab1d} 
\end{table}

\subsection{Material realizations}
\begin{figure}[tbp]
    \includegraphics[width=\columnwidth]{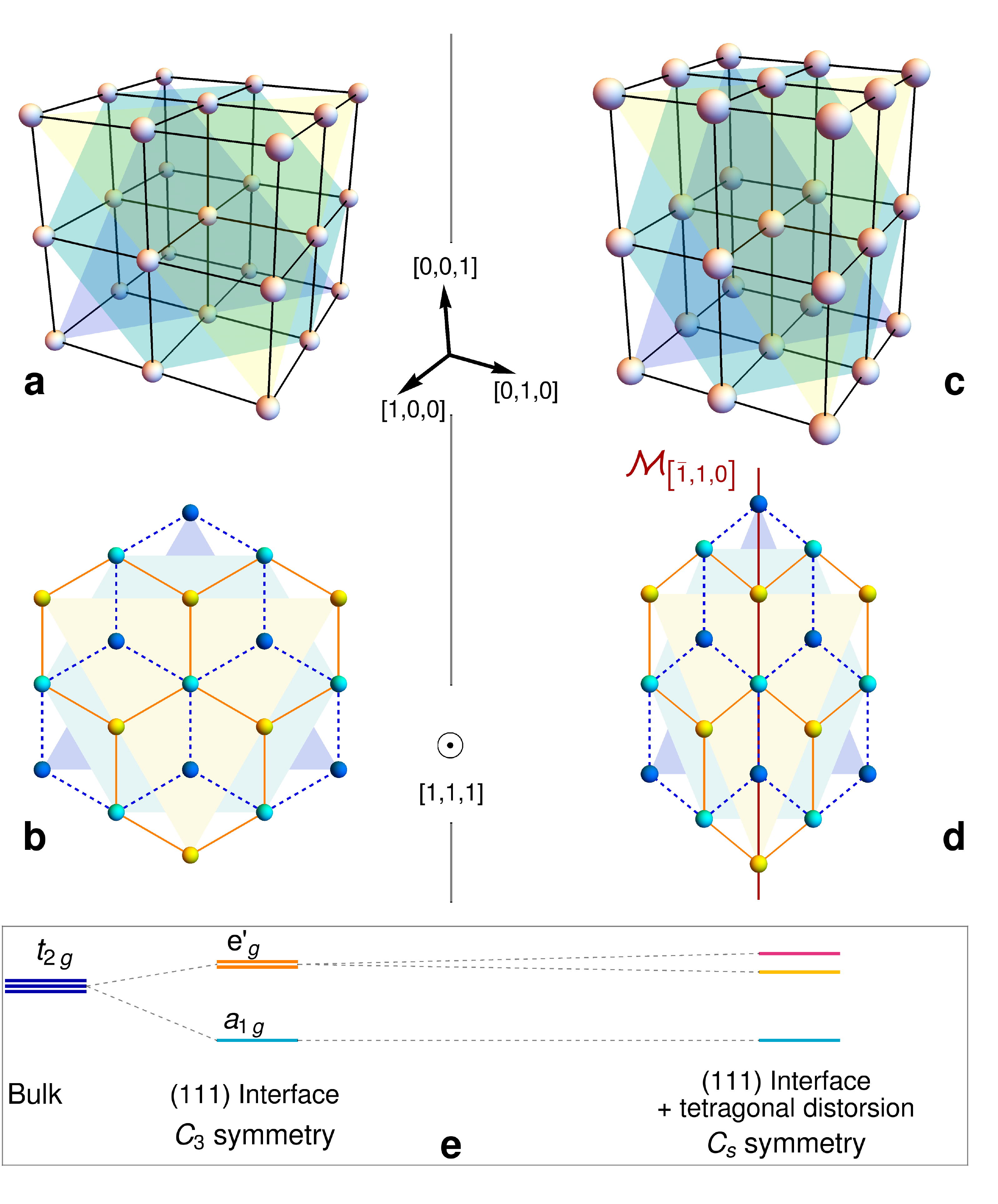}
    \caption{(a) Schematic representation of an ABO$_3$ pervoskite cubic unit cell displaying the three interlaced transition metal $[111]$ planes. (b) Corresponding top view along the $[111]$ crystallographic direction. We only show the $B$ transition metal atoms. (c) and (d) show the effect of a tetragonal distortion with the $[001]$ direction being the tetragonal axis. The distortion breaks the threefold rotation symmetry around the $[111]$ axis but leaves a residual mirror symmetry. (e) Evolution of the orbital states at the $\Gamma$ point of the BZ with quenched angular momentum.}
    \label{fig:fig2}
\end{figure}

Before analyzing the origin and physical consequence of the BC and its characteristic pinch points, we now introduce a material platform naturally equipped with orbital degrees of freedom and the required low crystalline symmetry: [111] interfaces of transition metal oxides hosting two-dimensional $d$ electron systems of $t_{2g}$ orbital character such as SrTiO$_3$~\cite{mon19,kha19}, KTaO$_3$~\cite{liu21}, and SrVO$_3$-based heterostructures. When compared to conventional semiconductor heterostructures, complex oxide interfaces consist of $d$ electrons with different symmetries, a key element in determining their many-body ground states that include, notably, unconventional superconductivity~\cite{rey07}. In the high-temperature cubic phases of these materials, the octahedral crystal field pins the low-energy physics to a degenerate $t_{2g}$ manifold, which spans an effective angular momentum one subspace, precisely as the $p$ orbitals discussed above~\footnote{We note that for the $t_{2g}$ orbitals the spin-orbit coupling has a minus sign due to the effective orbital moment projection from $L=2$ to the $L_{\text{eff}}=1$ in the $t_{2g}$ manifold. This overall sign however is not altering the results.}. The reduced symmetry at interfaces lift their energetic degeneracy and modify their orbital character. At the [111] interface the transition metal atoms form a stacked triangular lattice with three interlaced layers [see Fig.~\ref{fig:fig2}(a),(b)].  This results in a triangular planar crystal field that hybridizes the $\ket{xy}$, $\ket{xz}$ and $\ket{yz}$ orbitals to form an  $\ket{a_{1g}}= \left(\ket{xy} + \ket{xz} + \ket{yz}\right)/\sqrt{3}$ one-dimensional IRREP whereas the two states $\ket{e_{g \pm}^{\prime}}=\left(\ket{xy} + \omega^{\pm 1} \ket{xz} + \omega^{\pm 2} \ket{yz} \right)/\sqrt{3}$, with $\omega=\mathrm{e}^{2 \pi i/3}$, form the two-dimensional IRREP . 

The energetic ordering of the levels depends on the microscopic details of the interface. For example, at the (111)LaAlO$_3$/SrTiO$_3$ interface, x-ray absorption spectroscopy~\cite{del18} sets the $\ket{a_{1g}}$ state at lower energy [see Fig.~\ref{fig:fig2}(e)]. By further considering the structural inversion symmetry inherently present at the heterointerface, we thus formally reach the situation we discussed for the set of $p$ orbitals, be it for the trigonal symmetry that excludes any local concentrations of BC. However, and this is key, low-temperature phase transitions in oxides lower the crystal symmetry, often realising a tetragonal or orthorhombic phase with oxygen octahedra rotations and (anti)polar cation displacements. Let’s consider the paradigmatic case of SrTiO$_3$. A structural transition occurring at around 105~K, from the cubic phase to a tetragonal structure~\cite{fle68} [see Fig.~\ref{fig:fig2}(c)], breaks the threefold rotational symmetry leaving a single residual mirror line. Assuming the tetragonal axis to be along the $[001]$ direction, the surviving mirror symmetry at the $[111]$ interface corresponds to ${\mathcal M}_{[{\bar 1} 1 0]}$ [see Fig.~\ref{fig:fig2}(d)]. This structural distortion lifts the degeneracy of the $e_g^{\prime}$ doublet. The bonding and antibonding states $\ket{e_{g +}^{\prime}} \pm \ket{e_{g -}^{\prime}}$ have opposite mirror ${\mathcal M}_{[{\bar 1} 1 0]}$ eigenvalues and realize two distinct one-dimensional IRREP [see Fig.~\ref{fig:fig2}(e)]. 
SrTiO$_3$-based heterointerfaces undergo additional tetragonal to locally triclinic structural distortions at temperatures below $\simeq$~70~K which involves small displacements of the Sr atoms along the $[111]$ directions convoluted with TiO$_6$ oxygen-octahedron antiferrodistortive rotations~\cite{sal13}. In addition, below about 50~K, SrTiO$_3$ and KTaO$_3$ approach a ferroelectric instability that is accompanied by strong polar quantum fluctuations. This regime is characterized by a soft transverse phonon mode that involves off-center displacement of the Ti ions with respect to the surrounding octahedron of oxygen ions~\cite{ros13}, which, in the static limit, would correspond to a ferroelectric order parameter. This can potentially enhance the interorbital hybridization terms allowed in acentric crystalline environments, and thus boost the appearance of large BC concentrations.

\begin{figure*}[tbp]
    \includegraphics[width=\textwidth]{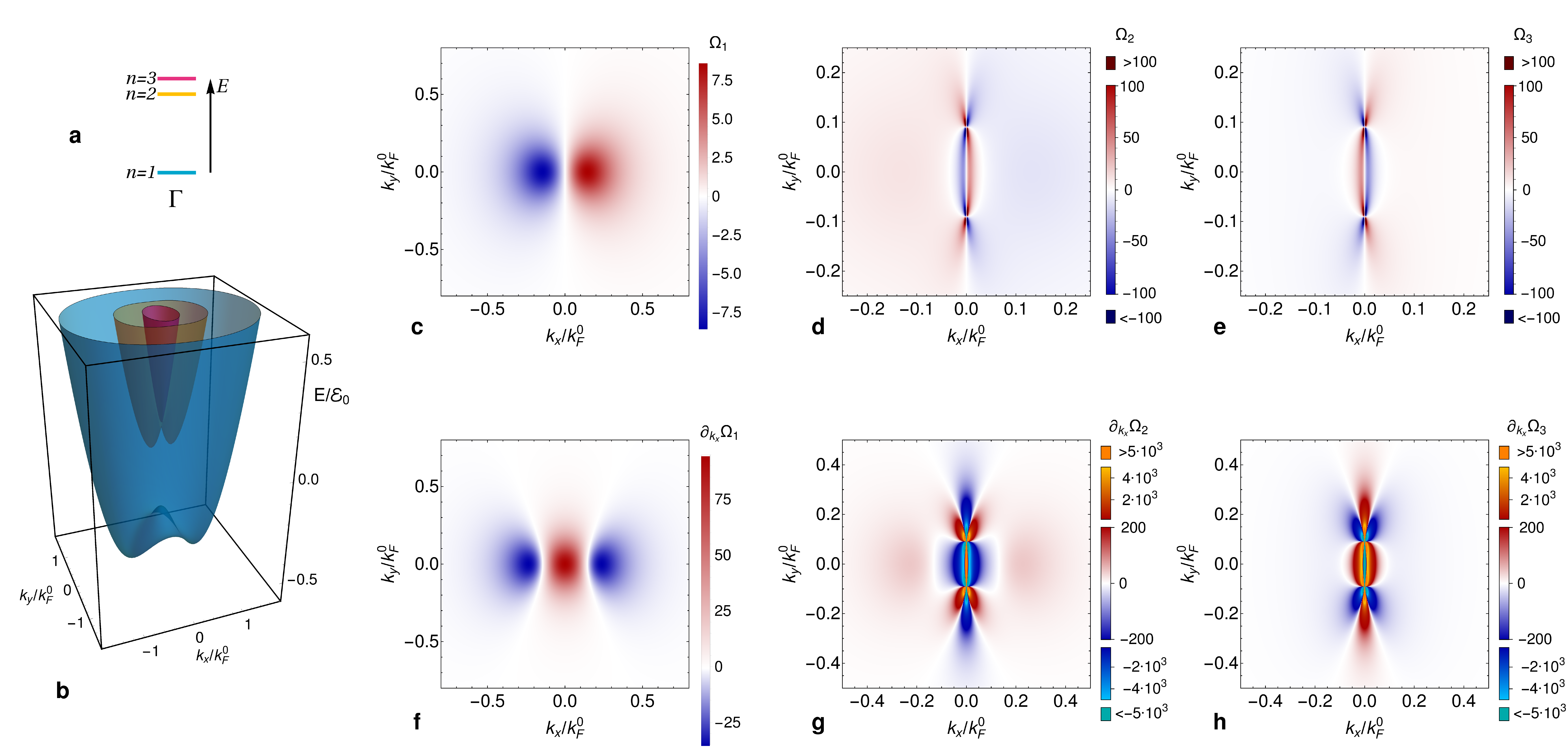}
    \caption{(a) Ordering of the crystal field split $t_{2g}$ ($p$) orbitals with their associated band index. (b) Energy spectrum of the model Eq.~\eqref{eq:Ham} obtained using the parameter set $\Delta=-0.2 {\cal E}_0$, $\Delta_m=0.01 {\cal E}_0$, $\alpha_R=\alpha_m=1.0 {\cal E}_0/k_F^0$. (c),(d),(e) show the ensuing band-resolved Berry curvature. (f), (g),(h) are the corresponding BC dipole densities $\partial_{k_x} \Omega$. Note that the presence of mirror symmetry guarantees that the orthogonal dipole density $\partial_{k_y} \Omega$ averages to zero.}
    \label{fig:fig3}
\end{figure*}

\subsection{Berry curvature dipole}
Having identified (111)-oriented oxide heterointerfaces as ideal material platforms, 
we next analyze the specific properties of the BC and its first moment. 
We first notice that in the case of a two-level spin system the local Berry curvature of the spin-split bands, if non-vanishing, is opposite.
Due to the concomitant presence of both spin-bands at each Fermi energy, 
the spin split bands cancel their respective local BC except for those momenta which are occupied by one spin band. 
In the $SU(3)$ system at hand, there is a similar sum rule stating that at each momentum ${\bf k}$ the BC of the three bands [c.f. Fig.~\ref{fig:fig3}(a)] sum to zero. However, and as mentioned above, the orbital bands are not subject to fermion multiplication theorems. 
In certain energy ranges a single orbital band is occupied  [c.f. Fig.~\ref{fig:fig3}(a,b)] and BC cancellations are not at work. 
There is also another essential difference between the BC associated to spin and orbital degrees of freedom. In general, the commutation and anticommutation relations of the $SU(N)$ Lie algebra define symmetric and antisymmetric structure constants, which, in turn, define the star and cross products of generic $SU(N)$ vectors~\cite{gra21}. 
Differently from an $SU(3)$ system spanning an angular momentum one subspace, in $SU(2)$ spin systems the symmetric structure constant vanishes identically. The ensuing absence of star products ${\bf b}_{\bf k} \star {\bf b}_{\bf k}$ precludes the appearance of BC with time-reversal symmetry as long as crystalline anisotropies are not taken into account [see Methods]. On the other hand, for $SU(3)$ the presence of all three purely imaginary Gell-Mann matrices $\Lambda_{2,5,7}$, together with the ``mass" terms $\Lambda_{3,8}$, 
is a sufficient condition to obtain time-reversal symmetric  BC concentrations even when accounting only for terms that are linear in momentum [see Methods]. 
This, however, strictly requires that  all rotation symmetries must be broken.

Next, we analyze the properties of the band resolved local BC 
starting
from the lowest energy band, which corresponds to the $\left(\ket{xy} + \ket{xz} + \ket{yz}\right)/\sqrt{3}$ state at $(111)$ LAO/STO heterointerfaces. 
Fig.~\ref{fig:fig3}(c) shows a characteristic BC profile. It displays two opposite poles centered on the $k_y=0$ line. 
These sources and sinks of BC are equidistant from the mirror symmetric $k_x=0$ line since the BC, as any genuine pseudoscalar, must be odd under vertical mirror symmetry operations, {\it i.e.} $\Omega(k_x, k_y)= - \Omega(-k_x, k_y)$. Note that the combination of time-reversal symmetry and vertical mirror implies that the BC will be even sending $k_y \rightarrow -k_y$, thus guaranteeing that, taken by themselves, the BC hot-spots will be centered around the $k_y=0$ line. 
Their finite $k_x$ values coincide with the points where the (direct) energy gap between the $n=1$ and the $n=2$ bands is minimized [see Fig.~\ref{fig:fig3}(a,b) and Supplemental Material], and thus the interorbital mixing is maximal. 
The properties of the BC are obviously reflected in the BCD local density $\partial_{k_x} \Omega(k_x, k_y)$: it possesses [see Fig.~\ref{fig:fig1}(f)] a positive area strongly localized at the center of the BZ that is neutralized by two mirror symmetric negative regions present at finite $k_x$. 
Let us next consider the Berry curvature profile arising from the two degenerate $e_{g}^{\prime}$ states that are split by the threefold rotation symmetry breaking. Fig.~\ref{fig:fig3}(d) shows the 
BC profile of the lowest energy band: it is entirely dominated by the BC pinch points induced by the mirror symmetry protected degeneracies on the $k_x=0$ line. The BC also displays a nodal ring around the pinch point,  and thus possesses a characteristic d-wave character around the singular point. 
This can be understood by constructing a ${\bf k \cdot p}$ theory around each of the two time-reversal related degeneracies. To do so, we first recall that the two bands deriving from the $e_g^{\prime}$ states have opposite ${\mathcal M}_x$ mirror eigenvalue along the full mirror line $k_x \equiv 0$ of the BZ. Close to the degeneracies, ${\mathcal M}_x$ can be therefore represented as $\sigma_z$. Under ${\mathcal M}_x$, $k_x \rightarrow -k_x$ whereas $k_y \rightarrow k_y$. Moreover, the Pauli matrices $\sigma_{x,y} \rightarrow -\sigma_{x,y}$. An effective two-band model close to the degeneracies must then have the following form at the leading order: 
\begin{equation}
\mathcal{H}_{eff}=v_x k_x~\sigma_x + \beta k_x~\delta k_y~\sigma_y+ v_y \delta k_y~\sigma_z, 
\end{equation}
where $\delta k_y$ is the momentum measured relatively to the mirror symmetry-protected degeneracy and 
we have neglected the quadratic term coupling to the identity $k^2 \sigma_0$ that does not affect the BC. Using the usual formulation of the BC for a two-band model [see the Methods section], it is possible to show that the Hamiltonian above is characterized by a zero-momentum pinch-point with two nodal lines [see the Supplemental Material] and d-wave character.
It is interesting to note that 
this
also implies that the ``effective" time-reversal symmetry inverting the sign of ${\bf k}$ around the pinch point is broken~\footnote{The effective time-reversal symmetry would imply that the BC should be an odd function of momentum}. Perhaps even more importantly, the d-wave character implies a 
very large BCD density in the immediate neighborhood of the pinch point [see Fig.~\ref{fig:fig3}(g)]. Similar properties are encountered when considering the highest energy band , with the difference that the pinch-point has an opposite angular dependence [see Fig.~\ref{fig:fig3}(e)] and consequently the BCD density has opposite sign [c.f. Fig.~\ref{fig:fig3}(h)].

\begin{figure*}[tbp]
    \includegraphics[width=\textwidth]{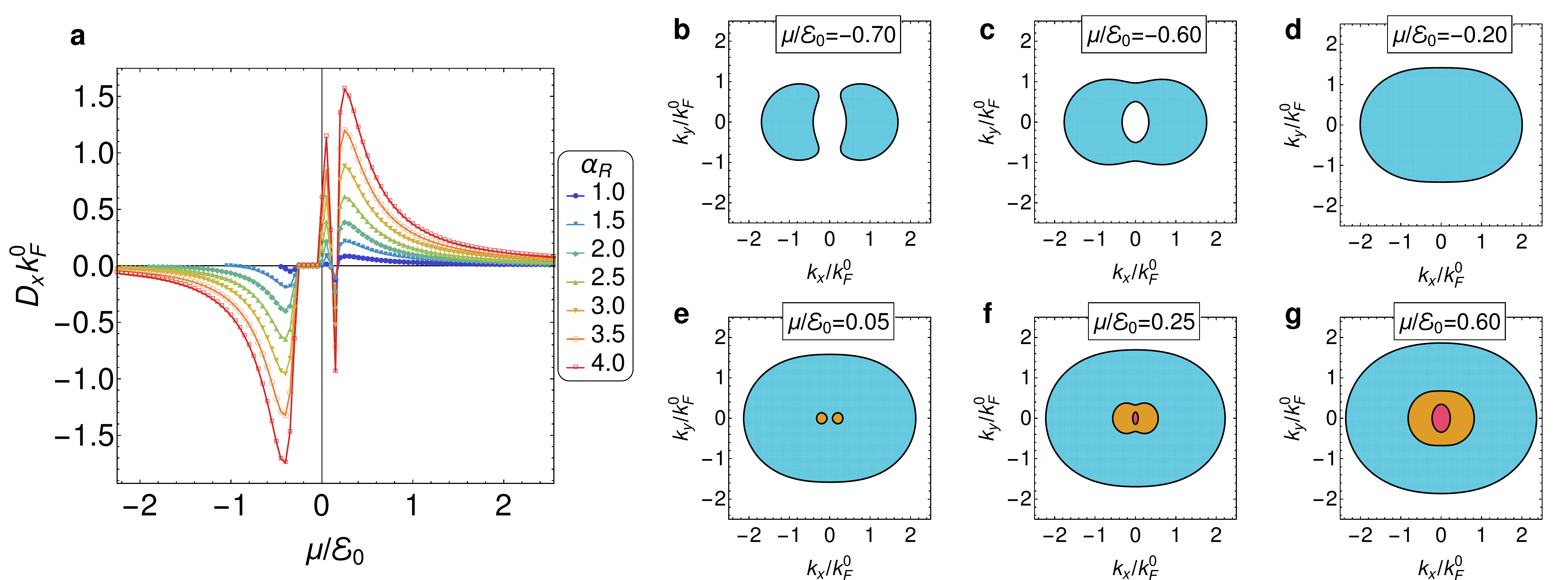}
    \caption{(a) Behavior of the Berry curvature dipole obtained by changing the chemical potential $\mu$ measured in units of ${\cal E}_0$. The different curves correspond to the different values of $\alpha_R$ measured in units of ${\cal E}_0/k_F^0$. The other model parameters have been instead fixed as $\Delta=-0.2 {\cal E}_0$, $\Delta_m=-0.01 {\cal E}_0 $, $\alpha_m=\alpha_R$. (b),(c),(d),(e),(f) display the Fermi lines and the band-resolved occupied regions in momentum space  for $\alpha_R= \alpha_m=1.5 {\cal E}_0/k_F^0$. }
    \label{fig:fig4}
\end{figure*}

Having the band-resolved BC and BCD density profiles in our hands, we finally discuss their characteristic fingerprints  in the BCD 
defined by $D_x= \int_{{\bf k}} \partial_{k_{x}} \Omega({\bf k}) f_0$, with $\int_{\bf k}=\int d^2 k / (2 \pi)^2$ and $f_0$ being the equilibrium Fermi-Dirac distribution function. 
By continuously sweeping the Fermi energy, we find that the BCD shows cusps and inflection points [see Fig.~\ref{fig:fig4}(a)] 
, which, as we now discuss,
are a direct consequence of Lifshitz transitions 
and their associated van Hove singularities [see Supplemental Material].
Starting from the bottom of the first band, the magnitude of the BCD continuously increases until it reaches a maximum where the dipole is larger than the inverse of the Fermi momentum of a 2DEG $1/k_F^{0}$ and thus gets an enhancement of three order of magnitudes with respect to a Rashba 2DEG~\cite{les22}. 
In this region, there are two distinct Fermi lines encircling electronic pockets at finite values of ${\bf k}$ [c.f. Fig.~\ref{fig:fig4}(b)], 
which subsequently merge on two disconnected regions in momentum space [c.f. Fig.~\ref{fig:fig4}(c)]. 
Since the states in the immediate vicinity of the center of the BZ are not occupied, the BCD is entirely dominated by the two mirror symmetric negative hot-spots of Fig.~\ref{fig:fig3}(f). By further increasing the chemical potential, the internal Fermi line collapses at the $\Gamma$ point and therefore a first Lifshitz transition occurs [c.f. Fig.~\ref{fig:fig4}(d)]. In this regime, the BCD has exponentially small values due to the fact that the strong positive BCD density area around the center of the BZ counteracts the mirror symmetric negative hot-spots. 
By further increasing the chemical potential, a second Lifshitz transition signals the occupation of the first $e_g$ band with two pockets centered around the $k_y=0$ line [see Fig.~\ref{fig:fig4}(e)]. This Lifshitz transition coincides with a rapid increase of the BCD due to the contribution coming from the local BCD density regions external to the BC nodal ring of Fig.~\ref{fig:fig3}(g). The subsequent sharp negative peak originates from a third Lifshitz transition in which the two electronic pockets of the second band merge, and almost concomitantly a tiny pocket of the third band centered around $\Gamma$ arises [see Fig.~\ref{fig:fig4}(f)]. By computing the band resolved BCD [see the Supplemental Material] one finds that it is this small pocket the cause of the negative sharp peak. 
For large enough chemical potentials, the BCD develops an additional peak corresponding to the fermiology of Fig.~\ref{fig:fig4}(g). This peak, which is again larger than $1/k_F^0$, can be understood by noticing that due to the BC local sum rule the momenta close to the center of the BZ do not contribute to the BCD. On the other hand, the regions external to the BC nodal ring are unoccupied by the third band and consequently have a net positive BCD local density. 
Thermal smearing can affect the strongly localized peaks at lower chemical potential but will not alter the presence of this broader peak. 
Note that the BCD gets amplified by increasing the interorbital mixing parameter $\alpha_R$ but retains similar properties [see Fig.~\ref{fig:fig4} and the Supplemental Material]. The strength of BC-mediated effects depends indeed on the ratio between the characteristic orbital Rashba energy $2 m \alpha_{R (m)}^2 / \hbar^2$ and the crystal field splittings $\Delta_{(m)}$.
The BCD
properties and values comparable to the Fermi wavelength are hence completely generic. 

Let us finally discuss the role of 
spin-orbit coupling. It can be included in our model Hamiltonian Eq.~\ref{eq:Ham} as
${\mathcal H}_{so}=\lambda_{so} \left(L_x \otimes \tau_x + L_y \otimes \tau_y + L_z \otimes \tau_z \right)$,
where $\lambda_{so}$ is the spin-orbit coupling strength, the $L=1$ angular momentum matrices correspond to the Gell-Mann matrices $\Lambda_2,\Lambda_5,\Lambda_7$, and the Pauli matrices $\tau_{x,y,z}$ act in spin space. 
Its effect can be analyzed using 
conventional (degenerate) perturbation theory. At the center of the Brillouin zone, ${\mathcal H}_{so}$ is completely inactive -- the eigenstates of the Hamiltonian Eq.~\ref{eq:Ham} are orbital eigenstates and the off-diagonal terms in orbital space $\Lambda_{2,5,7}$ cannot give any correction at first order in $\lambda_{so}$. 
The situation is different at finite values of momentum. The two spin-orbit free degenerate eigenstates are a superposition of the different orbitals (due to the orbital Rashba coupling). Therefore, the spin-orbit coupling term will lift their degeneracy resulting in a Rashba-like splitting of the bands.

In order to explore the consequence of this spin splitting  on the Berry curvature, let us denote with $\ket{\psi_0^{\uparrow}({\bf k})}$ and $\ket{\psi_0^{\downarrow}({\bf k})}$ the two spin-orbit free degenerate eigenstates at each value of the momentum. Note that $\ket{\psi_0}$ is a three-component spinor for the orbital degrees of freedom. 
When accounting perturbatively for spin-orbit coupling the eigenstates will be a superposition of the spin degenerate eigenstates and will generally read 
\begin{eqnarray}
\ket{\psi^{+}({\bf k})}=\cos{\theta({\bf k})} \mathrm{e}^{i \phi({\bf k})} \ket{\psi_0^{\uparrow}({\bf k})} + \sin{\theta({\bf k})}  \ket{\psi_0^{\downarrow}({\bf k})} \nonumber \\
\ket{\psi^{-}({\bf k})}=-\sin{\theta({\bf k})} \mathrm{e}^{i \phi({\bf k})} \ket{\psi_0^{\uparrow}({\bf k})} + \cos{\theta({\bf k})}  \ket{\psi_0^{\downarrow}({\bf k})} \nonumber 
\end{eqnarray}
Here, the momentum dependence of the phase $\phi$ and the angle $\theta$ is a ``by-product" of the orbital Rashba coupling: the effect of spin-orbit coupling, which is off-diagonal in orbital space, is modulated  by the momentum-dependent orbital content of the eigenstates $\ket{\psi_0^{\uparrow,\downarrow}({\bf k})}$.
The abelian Berry connection of the two spin-split states  
${\mathcal A}_{k_x,k_y}^{+,-}= \braket{\psi^{+,-}({\bf k}) | i \partial_{k_x,k_y}  \psi^{+,-}({\bf k})}$ will therefore contain two terms: the first one is the spin-independent Berry connection ${\mathcal A}^0_{k_x,k_y}=\braket{\psi_0({\bf k}) | i \partial_{k_x,k_y}  \psi_0({\bf k})}$; the second term is instead related to the derivatives of the phase $\phi$ and angle $\theta$. This Berry connection is opposite for the $+,-$ states and coincides with the Berry connection of a two-level spin system~\cite{xia10}. 
This also implies that the Berry curvature of a Kramers' pair of bands 
$\Omega^{+,-}({\bf k})=\Omega({\bf k}) \pm \Omega_{so}({\bf k}).$ 
The contribution of the Berry curvature $\Omega_{so}$ is opposite for the time-reversed partners and the net effect only comes from the difference between the Fermi lines of two partner bands. However, the purely orbital Berry curvature $\Omega({\bf k})$, which can be calculated directly from Eq.~\ref{eq:Ham}, sums up. The values of the BCD presented in Fig.~\ref{fig:fig4} are thus simply doubled in the presence of a weak but finite spin-orbit coupling.

\section{Conclusions}
In this study, we have shown an intrinsic pathway to design large concentrations of Berry curvature in time-reversal symmetric conditions making use only of the orbital angular momentum electrons acquire when bound to atomic nuclei. Such mechanism is different in nature with respect to that exploited in topological semimetals and narrow-gap semiconductors where the geometric properties of the electronic wavefunctions originate from the coupling between electron and hole excitations. 
The orbital design of Berry curvature is also inherently different from the time-reversal symmetric spin-orbit mechanism~\cite{he21,les22}
, which strongly relies on crystalline anisotropy terms. We have shown in fact that the Berry curvature triggered by orbital degrees of freedom features both hot-spots and singular pinch-points. 
Furthermore, due to the crystalline symmetry constraints the Berry curvature is naturally equipped with a non-vanishing Berry curvature dipole. These characteristics yield a boost of three orders of magnitude in the quantum non-linear Hall effect. 
In $(111)$ LaAlO$_3$-SrTiO$_3$ heterointerfaces where the characteristic Fermi wavevector $k_F^0 \simeq 1$~nm$^{-1}$, the Berry curvature dipole $D_x \simeq 1$nm. The corresponding non-linear Hall voltage can be evaluated using the relation~\cite{ma19,ho21} $V_{yxx}=e^3~\tau~D_x~|I_x|^2 / (2 \hbar^2 \sigma_{xx}^2 W)$, with the characteristic relaxation time $\tau \simeq 1$~pS and the longitudinal conductance $\sigma_{xx} \simeq 5$~mS. In a typical Hall bar of width $W \simeq 10 \mu$~m  sourced with a current $I_x \simeq 100~\mu$A, the non linear Hall voltage $V_{yxx} \simeq 2~\mu$V, which is compatible with the strong non-linear Hall signal experimentally detected~\cite{les22}.

The findings of our study carry a dramatic impact on the developing area of condensed matter physics dubbed orbitronics~\cite{go21}. Electrons in solids can carry information by exploiting either their intrinsic spin or their orbital angular momentum. Generation, detection and manipulation of information using the electron spin is at the basis of spintronics. The Berry curvature distribution we have unveiled in our study is expected to trigger also an orbital Hall effect, whose origin is rooted in the geometric properties of the electronic wavefunctions, and can be manipulated using the orbital degrees of freedom. This opens a number of possibilities for orbitronic devices. 
This is even more relevant considering that our findings can be applied to a wide class of materials whose electronic properties can be described with an effective $L=1$ orbital multiplet. These include other complex oxide heterointerfaces as well as spin-orbit free semiconductors where $p$-orbitals can be exploited. 
Since Dirac quasiparticles are not required in the orbital design of Berry curvature, it is possible to reach carrier densities large enough to potentially exploit electron-electron and electron-phonon interactions effects in the control of Berry curvature-mediated effects. For instance, orbital selective metal-insulator transitions can be used to switch on and off the electronic transport channels responsible for the Berry curvature and its dipole. We envision that this capability can be used to design orbitronic and electronic transistors relying on the geometry of the quantum wavefunctions.

\section*{Methods}
\subsection*{Representation of the Gell-Mann matrices in the symmetry groups}
\noindent
Apart from the identity matrix $\Lambda_0$, the eight Gell-Mann matrices can be defined as 
\begin{eqnarray*}
\Lambda_{1}=\begin{pmatrix}0 & 1 & 0\\
1 & 0 & 0\\
0 & 0 & 0
\end{pmatrix}, & \Lambda_{2}=\begin{pmatrix}0 & -i & 0\\
i & 0 & 0\\
0 & 0 & 0
\end{pmatrix},\\
\Lambda_{3}=\begin{pmatrix}1 & 0 & 0\\
0 & -1 & 0\\
0 & 0 & 0
\end{pmatrix}, & \Lambda_{4}=\begin{pmatrix}0 & 0 & 1\\
0 & 0 & 0\\
1 & 0 & 0
\end{pmatrix},\nonumber \\
\Lambda_{5}=\begin{pmatrix}0 & 0 & -i\\
0 & 0 & 0\\
i & 0 & 0
\end{pmatrix}, & \Lambda_{6}=\begin{pmatrix}0 & 0 & 0\\
0 & 0 & 1\\
0 & 1 & 0
\end{pmatrix},\nonumber \\
\Lambda_{7}=\begin{pmatrix}0 & 0 & 0\\
0 & 0 & -i\\
0 & i & 0
\end{pmatrix}, & \quad\Lambda_{8}=\begin{pmatrix}\tfrac{1}{\sqrt{3}} & 0 & 0\\
0 & \tfrac{1}{\sqrt{3}} & 0\\
0 & 0 & \tfrac{-2}{\sqrt{3}}
\end{pmatrix}.\nonumber 
\end{eqnarray*}
Let us now check the properties of these eight Gell-Mann matrices under time-reversal symmetry. Since we are considering electrons that are effectively spinless due to the $SU(2)$ spin symmetry, the time-reversal operator can be represented as ${\mathcal K}$. Hence, the three Gell-Mann matrices $\Lambda_2,\Lambda_5, \Lambda_7$ are odd under time-reversal, {\it i.e.} ${\mathcal T}^{-1} \Lambda_{2,5,7} {\mathcal T}=-\Lambda_{2,5,7}$, whereas the remaining matrices are even under time-reversal. Similarly, $\Lambda_{1,2,3,8}$ are even under the vertical mirror symmetry whereas $\Lambda_{4,5,6,7}$ are odd. Let us finally talk about the threefold rotational symmetry. Since the rotation symmetry operator ${\mathcal C}_3=\exp{\left[2 \pi i \Lambda_7 /3\right]}$, the transformation properties of the Gell-Mann matrices are determined by the commutation relations $\left[ \Lambda_7, \Lambda_i \right]$. The commutation relations are listed as follows:
\begin{eqnarray*}
\left[\Lambda_7, \Lambda_1 \right]=i \Lambda_4 & \hspace{.5cm} & \left[\Lambda_7, \Lambda_2 \right]=i \Lambda_5 \\
\left[\Lambda_7, \Lambda_4 \right]=-i \Lambda_1 & \hspace{.5cm} & \left[\Lambda_7, \Lambda_5 \right]=-i \Lambda_2 \\
\left[\Lambda_7, \Lambda_6 \right]= 2 i \left(\frac{\Lambda_3}{2} - \frac{\sqrt{3}}{2} \Lambda_8 \right) & \hspace{.5cm} & \left[\Lambda_7 , \frac{\Lambda_3}{2} - \frac{\sqrt{3}}{2} \Lambda_8 \right] =  - 2 i \Lambda_6 \\ 
\left[\Lambda_7, \Lambda_3 + \frac{\Lambda_8}{\sqrt{3}} \right]=0 & & 
\end{eqnarray*}
The results above indicate that the three pairs of operators $\left\{\Lambda_{1}, \Lambda_{4} \right\}$, $\left\{\Lambda_2 , \Lambda_5 \right\}$, and $ \left\{ \Lambda_6, \frac{\Lambda_3}{2} - \frac{\sqrt{3}}{2} \Lambda_8 \right\}$ behave as vector under the threefold rotation symmetry and therefore form two-dimensional IRREPS. 

\subsection*{Berry curvature of $SU(2)$ and $SU(3)$ systems}

For SU(2) systems, a generic Hamiltonian can be written in terms of Pauli matrices $\sigma_i$  as ${\mathcal H}({\bf k})=d_0({\bf k})\sigma_0+ \bf{d}({\bf k})\cdot \bm{\sigma}$, where $\sigma_0$ is  the $2\times2$ identity matrix and the Pauli matrix vector $\bm{\sigma}=(\sigma_x,\sigma_y\sigma_z)$. The Berry curvature can be expressed in terms of $\bf{d}$ vector
\begin{equation}
\label{berrySU2}
\Omega_{\pm}(\mathbf{k})= \mp \frac{1}{2|{\bf d}({\bf k})|^3}{\bf d({\bf k})}\cdot [\partial_{k_x} \mathbf{d}({\bf k}) \times \partial_{k_y} \mathbf{d}({\bf k})]\;.
\end{equation}
For SU(3) system, we can proceed 
analogously  using the Gell-Mann matrices introduced above. The Hamiltonian  of a system described by three electronic degrees of freedom in a $3\times3$ manifold can be written as ${\mathcal H}({\bf k})=b_0({\bf k})\Lambda_0+ {\bf b}({\bf k})\cdot \bm{\Lambda}$, where $b_0({\bf k})$ is a scalar and ${\bf b}({\bf k})$ is an eight dimensional vector. 
The Gell-Mann matrices satisfy an algebra which is a generalization of the SU(2). In particular, we have that 
\begin{equation}
\Lambda_{a} \Lambda_{b}=\tfrac{2}{3}\delta_{a b}+ \left(d_{abc}+ i f_{abc}\right)\Lambda_{c}, 
\label{mult}
\end{equation}
where repeated indices are summed over. 
In the equation above, we have introduced the antisymmetric and symmetric structure factors of SU(3) that are defined respectively as 
\begin{eqnarray}
\nonumber
f_{abc}=-\frac{i}{4}{\rm Tr}([\Lambda_a,\Lambda_b]\Lambda_c)\;, \quad
d_{abc}=\frac{1}{4}{\rm Tr}(\{\Lambda_a,\Lambda_b\}\Lambda_c).
\end{eqnarray}
From these one defines three bilinear operations of SU(3) vectors:
the dot (scalar) product $\mathbf{v}\cdot \mathbf{w} = v_{a}w_{a}$, the cross product $(\mathbf{v}\times \mathbf{w})_a = f_{abc}v_{b}w_{c}$,  and 
the star product $(\mathbf{v}\star \mathbf{w})_a = d_{abc}v_{b}w_{c}$.
The star product is a symmetric vector product which does not play any role for SU(2) since $d_{abc}=0$. 
Moreover, the band-resolved Berry curvature is given by~\cite{bar12,gra21}:
\begin{eqnarray}
\label{BerrySU3}
\Omega_n ({\bf k}) &=& -4\frac{(\gamma_{{\bf k},n} \mathbf{b}_{{\bf k}} + \mathbf{b}_{{\bf k}}\star\mathbf{b}_{{\bf k}})}
{(3\gamma^2_{{\bf k},n} -|\mathbf{b}_{{\bf k}}|^2)^3} \cdot 
\left\{ [\gamma_{{\bf k},n}  \partial_{k_x}\mathbf{b}_{{\bf k}} +  \right. \\ \nonumber
 & +& \left. \partial_{k_x} (\mathbf{b}_{{\bf k}} \star \mathbf{b}_{{\bf k}})] \times 
[\gamma_{{\bf k},n}  \partial_{k_y}\mathbf{b}_{{\bf k}} +\partial_{k_y} (\mathbf{b}_{{\bf k}} \star \mathbf{b}_{{\bf k}})] \right\} 
\end{eqnarray}
where we introduced $\gamma_{{\bf k},n}  = \frac{2}{\sqrt{3}} |\mathbf{b}_{{\bf k}}|\cos\left(\theta_{{\bf k}}+\tfrac{2\pi}{3}n\right)$, 
$\theta_{{\bf k}} =  \tfrac{1}{3}\arccos\left[\frac{\sqrt{3}\mathbf{b}_{\bf k}\cdot
\left( \mathbf{b}_{\bf k}\star \mathbf{b}_{{\bf k}}\right)}{|\mathbf{b}_{\bf k}|^{3}}\right]$, and 
$\mathbf{b}_{{\bf k}}$ is a shorthand for ${\bf b}({\bf k})$. 
Generally speaking, the Berry curvature in Eq.~(7) can be split in two contributions as $\Omega_n ({\bf k}) = \Omega^{(0)}_n({\bf k}) + \Omega^{(\star)}_n({\bf k})$, with  $\Omega^{(0)}_n({\bf k})= -4\frac{(\gamma_{{\bf k},n})^3}{(3\gamma^2_{{\bf k},n} -|\mathbf{b}_{{\bf k}}|^2)^3} \mathbf{b}_{{\bf k}} \cdot  [\partial_{k_x} \mathbf{b}_{{\bf k}} \times \partial_{k_y} \mathbf{b}_{{\bf k}}] $  that strongly resembles the BC expression for SU(2) systems of Eq.~(6).  
In trigonal systems described by the 
effective Hamiltonian in Eq.~\eqref{eq:H0} we have that for any momentum ${\bf k}$ and for any value of the parameters, the ${\bf b}$ vector associated with the Hamiltonian is such that $(\gamma_{{\bf k},n} \mathbf{b}_{{\bf k}} + \mathbf{b}_{{\bf k}}\star\mathbf{b}_{{\bf k}}) $ is always orthogonal to the vector in the curly braces in the expression of the Berry curvature Eq.~\eqref{BerrySU3}. Hence $\Omega_n({\bf k})=0$.
On the contrary, assuming a ${\mathcal C}_s$ point-group symmetry the effective Hamiltonian of Eq.~\eqref{eq:Ham} defines 
$\mathbf{b}_{{\bf k}} = \left(0,\alpha_R k_y,\Delta+\frac12\Delta_m,0,\alpha_R k_x, 0,\alpha_m k_x, 
\frac{\Delta}{\sqrt{3}}-\frac{\sqrt{3}}{2}\Delta_m\right)$,
for which we get that $\Omega^{(0)}_n({\bf k})=0$, and the BC is substantially given by  $\Omega^{(\star)}_n({\bf k})$. 
In other words, 
the terms obtained by doing the star product, $\mathbf{b}_{{\bf k}}\star\mathbf{b}_{{\bf k}}$
are those that yield the non-zero BC.
We point out that the BC is proportional to the combination of parameters $\alpha_R^2~\alpha_m~\left(2 \Delta + \Delta_m\right)$. A non-vanishing BC can be thus obtained even in the absence of the Gell-Mann matrix $\Lambda_8$. This, on the other hand, would correspond to values of the crystal field splitting $\Delta_m=2 \Delta /3$ implying a very strong distortion of the crystal from the trigonal arrangement. The presence of the constant term $\propto \Lambda_8$ is thus essential to describe systems with a parent high-temperature trigonal crystal structure. 
 
\subsection*{Calculation of the Berry curvature dipole}

The first moment of the Berry curvature, the Berry curvature dipole, for each energy band $n$ is given by 
$D_{x,n}=\int \frac{d^2 k}{(2\pi)^2} \partial_{k_x} \Omega_n({\bf k}) f_0 ({\bf k})$, 
where $f_0 ({\bf k})$ is the equilibrium Fermi-Dirac distribution function. 
At zero temperature, this expression can be rewritten as a line integral
over the Fermi line
\begin{equation}
\label{lineI}
D_{x,n}=\int \frac{d^2 k}{(2\pi)^2} \Omega_n({\bf k}) \frac{\partial E_n}{\partial k_x} \delta(E_n -\mu)\;,
\end{equation}
where $E_n=E_n({\bf k})\; (n=1,2,3)$ are the energy bands and $\mu$ is the chemical potential. We have used the latter expression \eqref{lineI} to evaluate the BCD, where $D_x= \sum_{n=1}^3 D_{x,n}$.
\vspace{0.4cm}

\noindent 
\textbf{Competing interests:} The authors declare no competing financial or non-financial interests. 
\\
\textbf{Author contributions:} C.O. conceived and supervised the project. M.T.M. performed the computations with help from C.N. and M.C. All authors participated in the analysis of the results and in the writing of the manuscript.

%

\end{document}


\title{Supplemental Material for: \\``Orbital design of Berry curvature: pinch points and giant dipoles induced by crystal fields"}

\author{Maria Teresa Mercaldo}
\affiliation{Dipartimento di Fisica ``E. R. Caianiello", Universit\`a di Salerno, IT-84084 Fisciano (SA), Italy}

\author{Canio Noce}
\affiliation{Dipartimento di Fisica ``E. R. Caianiello", Universit\`a di Salerno, IT-84084 Fisciano (SA), Italy}
\affiliation{SPIN-CNR, IT-84084 Fisciano (SA), Italy}

\author{Andrea D. Caviglia}
\affiliation{Department of Quantum Matter Physics, University of Geneva, 24 Quai Ernest Ansermet, CH-1211 Geneva, Switzerland}

\author{Mario Cuoco}
\affiliation{SPIN-CNR, IT-84084 Fisciano (SA), Italy}
\affiliation{Dipartimento di Fisica ``E. R. Caianiello", Universit\`a di Salerno, IT-84084 Fisciano (SA), Italy}
\author{Carmine Ortix}
\affiliation{Dipartimento di Fisica ``E. R. Caianiello", Universit\`a di Salerno, IT-84084 Fisciano (SA), Italy}

\maketitle


\section{Berry curvature properties}

\begin{center}
\begin{figure*}[h!]
\includegraphics[width=0.9\textwidth]{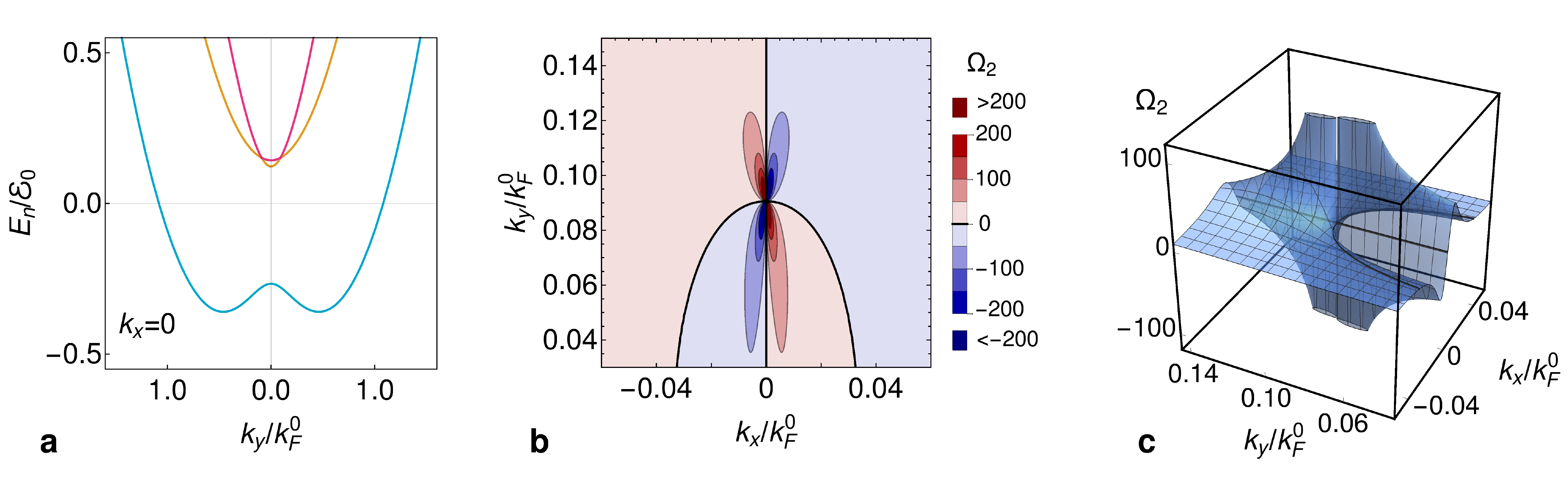} 
\protect\caption{{\bf a} Energy dispersion of the orbital model along the mirror symmetric line of the BZ $k_x=0$. At finite $k_y$ there are two mirror symmetry-protected crossings {\bf b} Contour plot and {\bf c} tridimensional plot of the band-resolved Berry curvature corresponding to the second band $\Omega_2({\bf k})$ in the region close to the mirror-symmetry protected crossing where the BC pinch point is formed. We have used the same set of parameters of Fig.~3 in the main article, i.e. $\Delta=-0.2 {\cal E}_0$, $\Delta_m=-0.01 {\cal E}_0$, and   $\alpha_R=\alpha_m=1.0 {\cal E}_0/k_F^0$. }
\label{figSM1}
\end{figure*}
\end{center}

\begin{center}
\begin{figure*}[h!]
    \includegraphics[width=.3\textwidth]{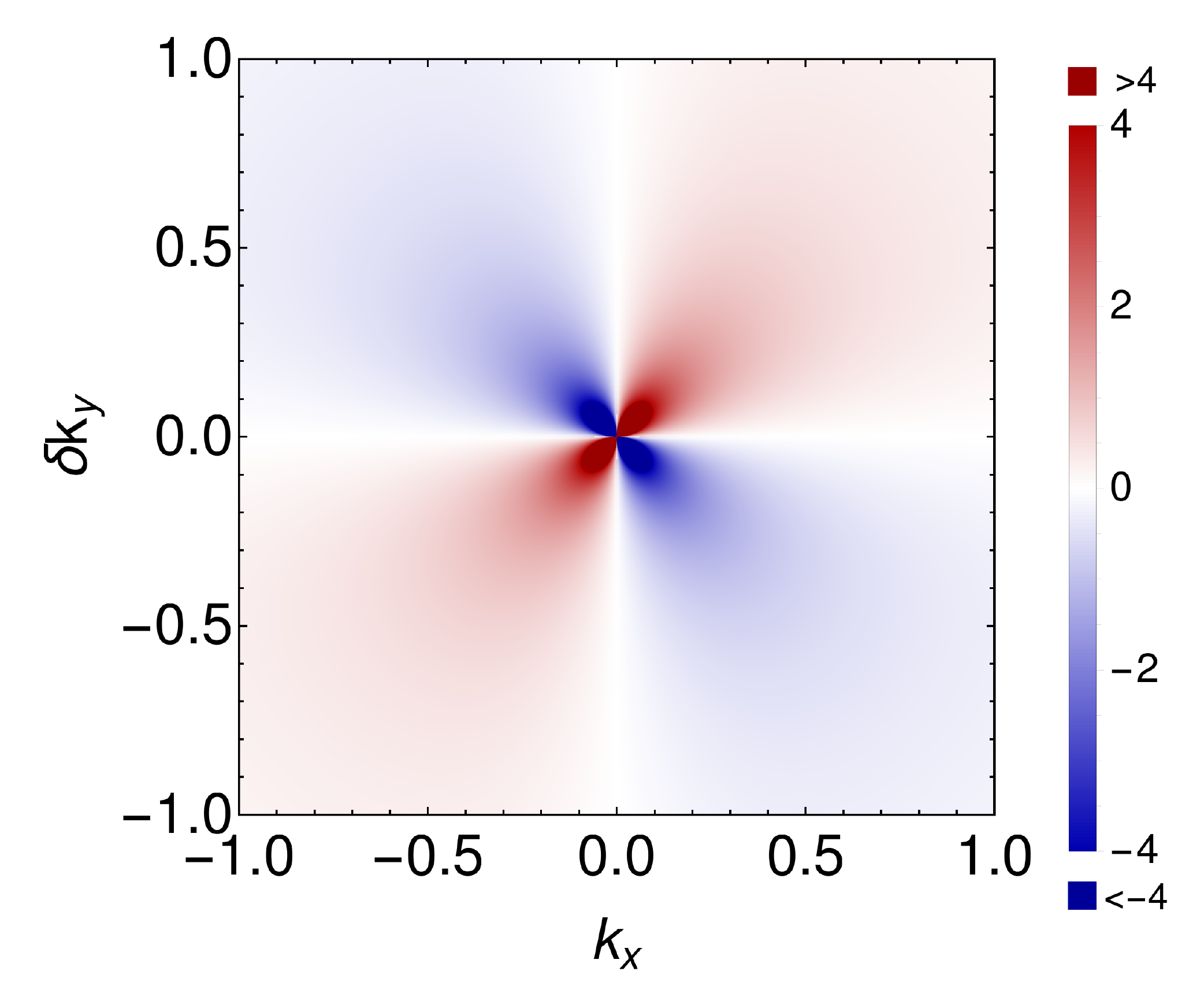}
    \caption{Density plot of the Berry curvature corresponding to the effective two-band model of Eq.~\ref{eq:hameffective} with a pinch point at zero momentum assuming $v_x=v_y=\beta$.}
    \label{figSMrev1}
\end{figure*}
\end{center}

In Fig.~\ref{figSM1} we show the connection between the presence of mirror-symmetry protected crossings and the Berry curvature pinch-points. 
This is further highlighted in Fig.~\ref{figSMrev1} where we display the Berry curvature distribution associated with the effective ${\bf k \cdot p}$ theory model close to a mirror symmetry protected degeneracy and reading (see the main text) 
\begin{equation}
{\mathcal H}_{eff}=v_x k_x \sigma_x + \beta k_x \delta k_y \sigma_y + v_y \delta k_y \sigma_z, 
\label{eq:hameffective}
\end{equation}
where $\delta k_y$ is the momentum measured relatively to the mirror symmetry-protected degeneracy.

\begin{center}
\begin{figure*}[h!]
\includegraphics[width=0.98\textwidth]{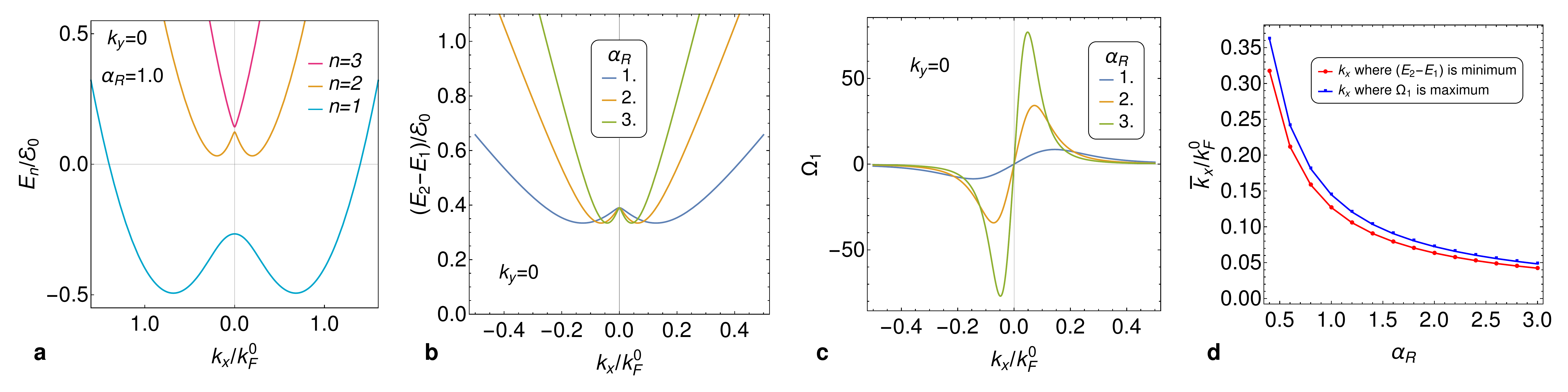} 
\protect\caption{(a) Energy dispersion along the $k_y=0$ line for $\Delta=-0.2 {\cal E}_0$, $\Delta_m=-0.01 {\cal E}_0$,  $\alpha_R=\alpha_m=1.0 {\cal E}_0/k_F^0$.  (b)  The corresponding behavior of the energy difference between the first two bands  $E_2-E_1$. We show also other values of the interorbital mixing parameter $\alpha_R$.(c) The local behavior of the Berry curvature of the first band $\Omega_1$ for different values of $\alpha_R$ shows that large enhancements of the Berry curvature, {\it i.e.} the hot-spots, are strongly related to the $E_2-E_1$ minima. This is further shown in (d) where we compare the location in ${\bf k}$ space of the BC hot-spots and the $E_2-E_1$ minima.}
\label{figSM2}
\end{figure*}
\end{center}

\begin{center}
\begin{figure*}[h!]
\includegraphics[width=0.3\textwidth]{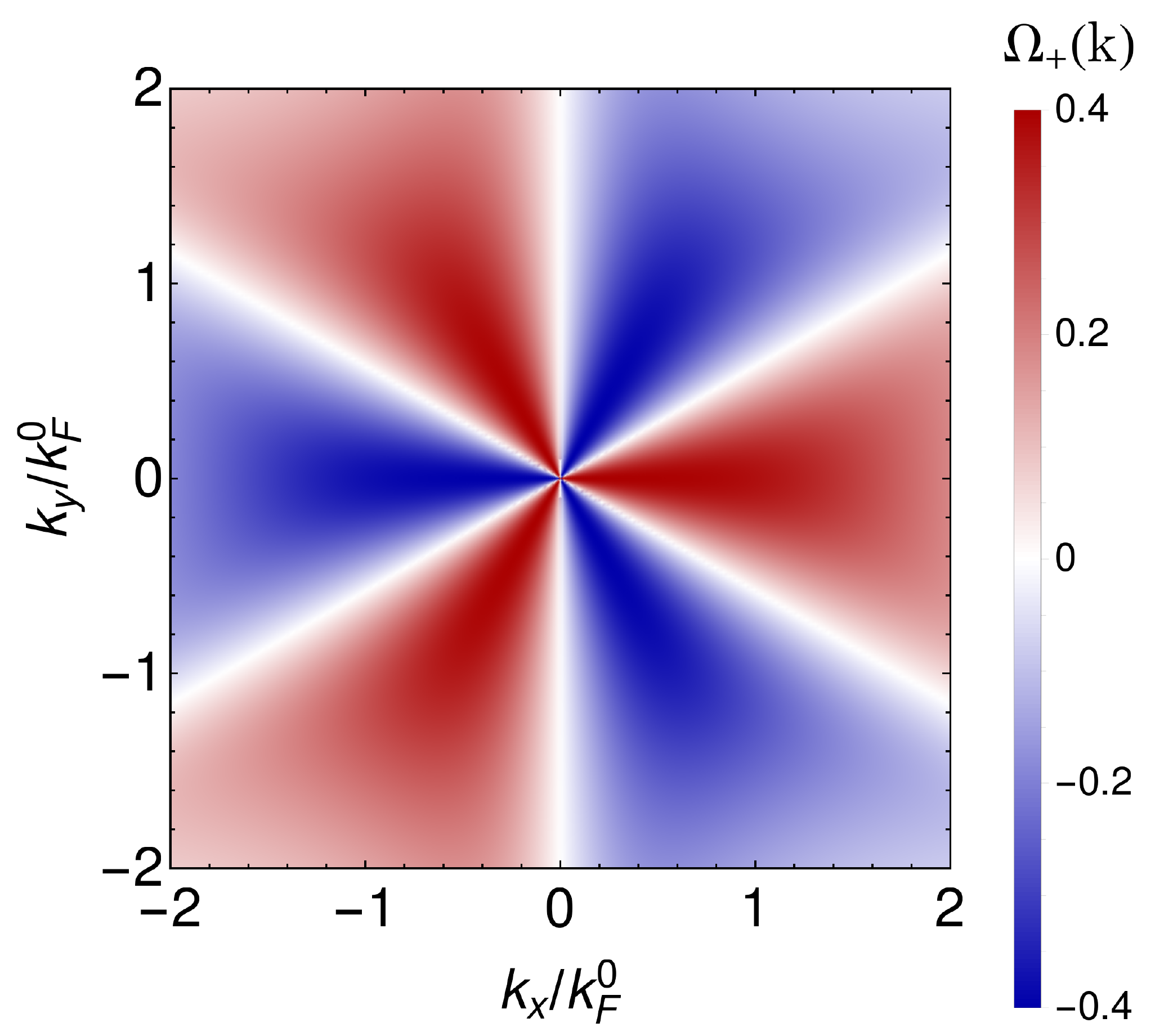} 
\protect\caption{Density plot of the Berry curvature for a spin-orbit coupled electron gas with warping terms cubic in momentum with effective Hamiltonian Eq.~\ref{eq:h2deg}. Here we have chosen $\alpha_{SR}=1$ and $\lambda=0.2$. }
\label{figSMrev2}
\end{figure*}
\end{center}

Fig.~\ref{figSM2} illustrates the relation between the occurrence of the Berry curvature hot-spots and the energetic distance between the $\ket{a_{1g}}$ and the crystal field split $\ket{e_g^{\prime}}$ state. 
Finally, in Fig.~\ref{figSMrev2} we show for comparison the Berry curvature of a Rashba two-dimensional electron gas with crystalline anisotropy terms that are cubic in momentum. The corresponding Hamiltonian reads: 
\begin{equation}
{\mathcal H}_{2DEG}=\dfrac{\hbar^2 k^2}{2 m} \sigma_0 + \alpha_{SR} \left(k_x \sigma_y - k_y \sigma_x \right) + \dfrac{\lambda}{2} \left(k_{+}^3 + k_{-}^3 \right) \sigma_z 
\label{eq:h2deg}
\end{equation}
where $k_{\pm}=k_x \pm i k_y$.

\section{Relation between van Hove singularities and properties of the Berry curvature dipole}

\begin{center}
\begin{figure*}[h!]
\includegraphics[width=0.4\textwidth]{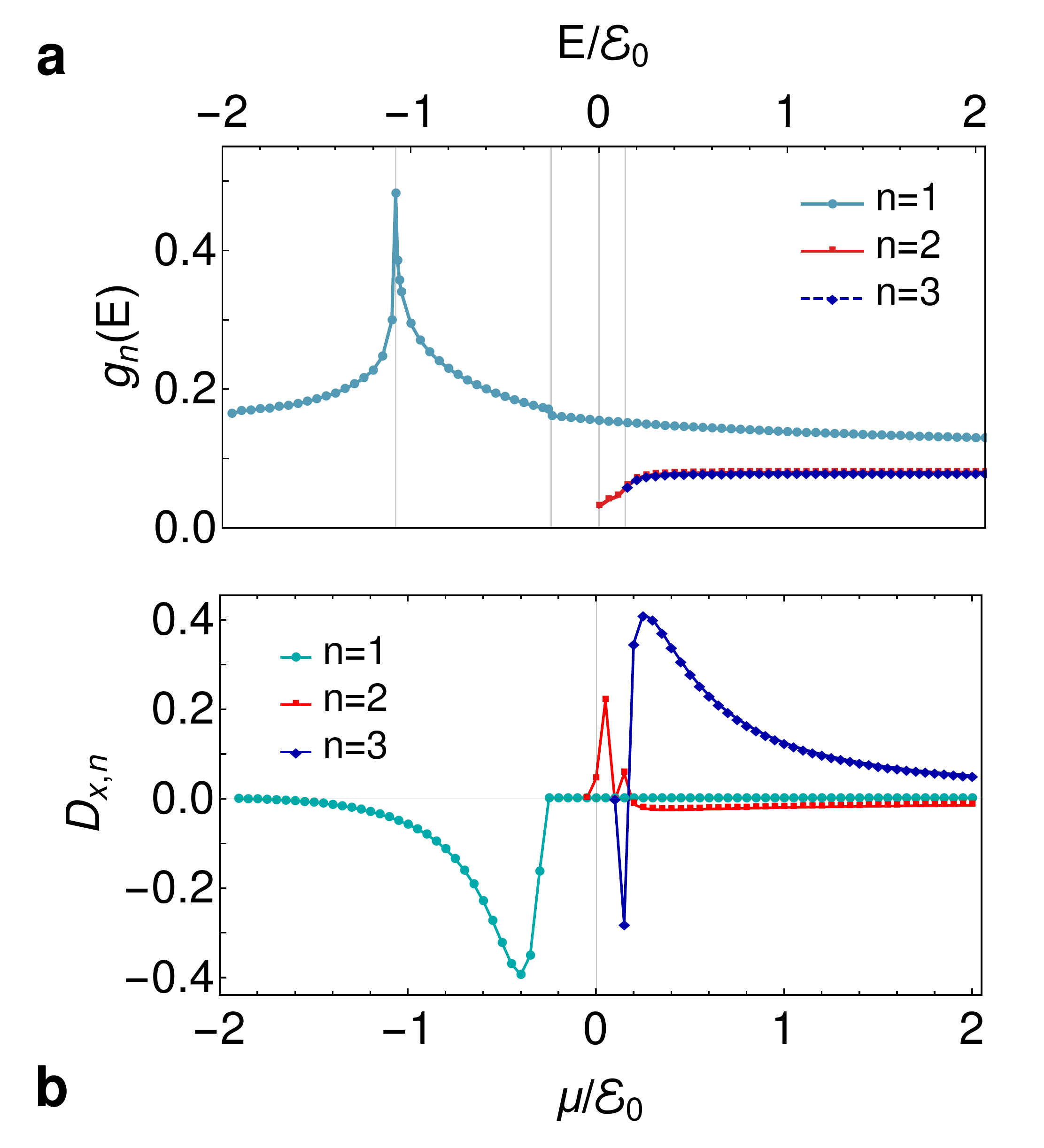} 
\protect\caption{(a) Band-resolved density of states  $g_n(E) $ and (b) Berry curvature dipole for 
  $\Delta=-0.2{\cal{E}}_{0}$, $\Delta_m=0.01 {\cal{E}}_{0} $, 
  $\alpha_{OR} = 2.0 {\cal{E}}_{0} / k_F^0$, $\alpha_m=\alpha_{OR}$. In panel (a) the vertical lines indicate the various Lifshitz transitions.}
\label{figSMrev3}
\end{figure*}
\end{center}

\begin{center}
\begin{figure*}[h!]
\includegraphics[width=0.99\textwidth]{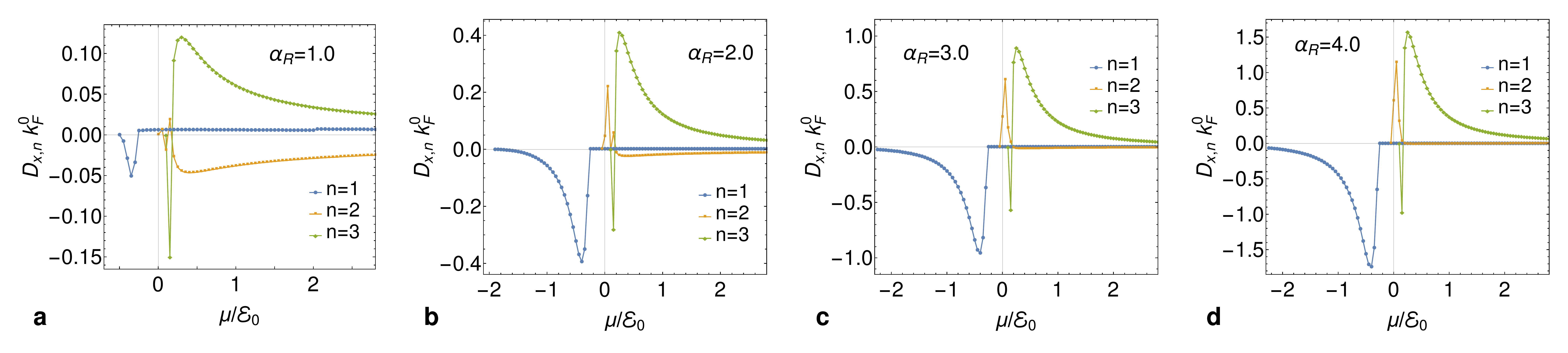} 
\protect\caption{The band-resolved Berry curvature dipole for different values of the interorbital mixing parameter $\alpha_R$.  
The other parameters are the same as in the main text.}
\label{figSM3}
\end{figure*}
\end{center}

As mentioned in the main text, the cusps and inflection points in the Berry curvature dipole are direct consequence of  Lifshitz transitions and their associated van Hove singularities. We recall that the generic sequence of Lifshitz transition in our model is as follows (see also Fig.4 of the main text).  First, there are two ``small" electron pockets centered at finite momentum that then merge to create a single electron pocket (centered at $\Gamma$) delimited by two ``concentric" Fermi lines. Note that this Fermiology is strongly reminiscent of a conventional isotropic Rashba two-dimensional electron gas in the low-density regime. The internal Fermi line then collapses at zero momentum and a single Fermi line survives. Such intraband Lifshitz transitions are then followed by the interband Lifshitz transition due to the appearance of two pockets of the second band that then merge with the almost concomitant appearance of a tiny pocket of the third (highest-in-energy) band. As shown in Fig.~\ref{figSMrev3} all these Lifshitz transitions are characterized by van Hove singularities. Indeed, at all Lifshitz transitions there is a divergence in the slope of the density of states. We point out that at the first intraband Lifshitz transition, the singularities due to the saddle points of the energy levels are not integrable and a divergence in the density of states occurs. This is because the saddle points are distributed over lines of the Brillouin zone -- similarly to the zero-energy van Hove divergence of a Rashba two-dimensional electron gas. Importantly, while the cusps and inflection points of the Berry curvature dipole can be all related to Lifshitz transitions (and hence van Hove singularities), the opposite does not hold. The direct comparison between Berry curvature dipole and density of states [see Fig.~\ref{figSMrev3}] indeed shows that the divergence of the density of states is accompanied by a featureless Berry curvature dipole. Note that the features of the Berry curvature dipole are preserved by changing the interorbital mixing parameter $\alpha_R$ as shown in Fig.~\ref{figSM3}.

\section{Effect of a constant $\Lambda_1$ term}
As mentioned in the main text, reducing the point-group symmetry from ${\mathcal C}_{3v}$ to ${\mathcal C}_s$ makes a constant term in $\Lambda_1$ symmetry allowed. This term is not present in our model Hamiltonian. It is important to note that such term yields a mixing between two states with equal ${\mathcal M}_x$ mirror eigenvalue (for $p$ orbitals these correspond to the $\ket{p_z}$ and $\ket{p_y}$ states). Consequently, a constant $\Lambda_1$ term lifts the energy degeneracy at the center of the BZ of the $e_g^{\prime}$ ($\ket{p_x}$, $\ket{p_y}$) states precisely as the constant term in $\Lambda_3/2 - \sqrt{3}\Lambda_8/2$ does. However, there is an important difference. The constant $\Lambda_3/2 - \sqrt{3}\Lambda_8/2$ term, appearing in Eq.5 of the main text, corresponds to a direct splitting between the two states of the doublet. On the contrary, the splitting due to the constant $\Lambda_1$ term is suppressed by the energy difference -- caused by the trigonal crystal field -- between the $a_{1g}$ ($\ket{p_z}$) state and the $e_g^{\prime}$ doublet. Since the strengths $\Delta_1,\Delta_m$ of the constant terms in $\Lambda_1$ and $\Lambda_3/2 - \sqrt{3}\Lambda_8/2$ are expected to have the same order of magnitude (both terms are proportional to the degree of lattice distortion from ${\mathcal C}_{3v}$) one finds that the constant term $\propto \Lambda_1$ has a much weaker effect on the electronic properties: the band structure and the behavior of the Berry curvature are essentially determined by $\Delta_m$. 
In particular, neither the existence of Berry curvature hot-spots nor the pinch points are affected by $\Delta_1$, in this regime.

\begin{figure*}[tbp]
\begin{centering}
    \includegraphics[width=0.3\textwidth]{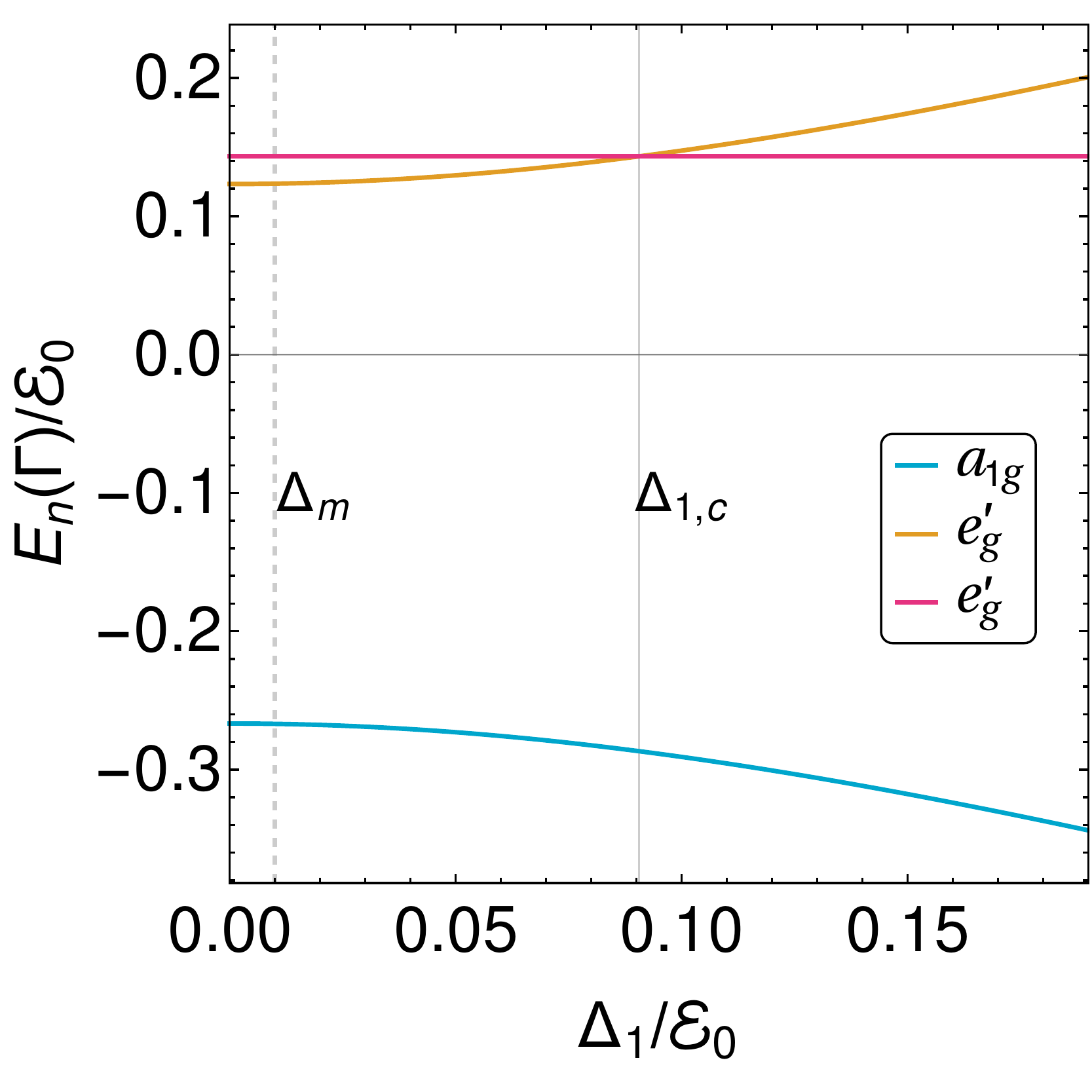}
    \caption{Behavior of the energy levels of the $a_{1g}$ and $e_{g}^{\prime}$ doublet (split by the crystal field $\Delta_m$) as a function of the $\Delta_1$ mixing term. The critical value $\Delta_{1,c}$ where a level crossing is obtained is much larger than $\Delta_m=0.01 {\cal E}_0$. The trigonal crystal field splitting $\Delta=-0.2 {\cal E}_0$ as in the main text.}
    \label{figSMrev4}
\end{centering}
\end{figure*}

To 
illustrate
this point, 
we show in Fig.~\ref{figSMrev4} the behavior of the energies (at the $\Gamma$ point) of the three $a_{1g}$ and $e_g^{\prime}$ states by increasing the strength of the constant $\Lambda_1$ term at fixed value of the crystal field splittings $\Delta,\Delta_m$. One finds that a non-zero $\Delta_1$ is able to change the level ordering -- and thus have a dramatic effect on the band structure -- only for $\Delta_1 \gg \Delta_m$.

\begin{figure*}[h!]
\begin{centering}
    \includegraphics[width=0.6\textwidth]{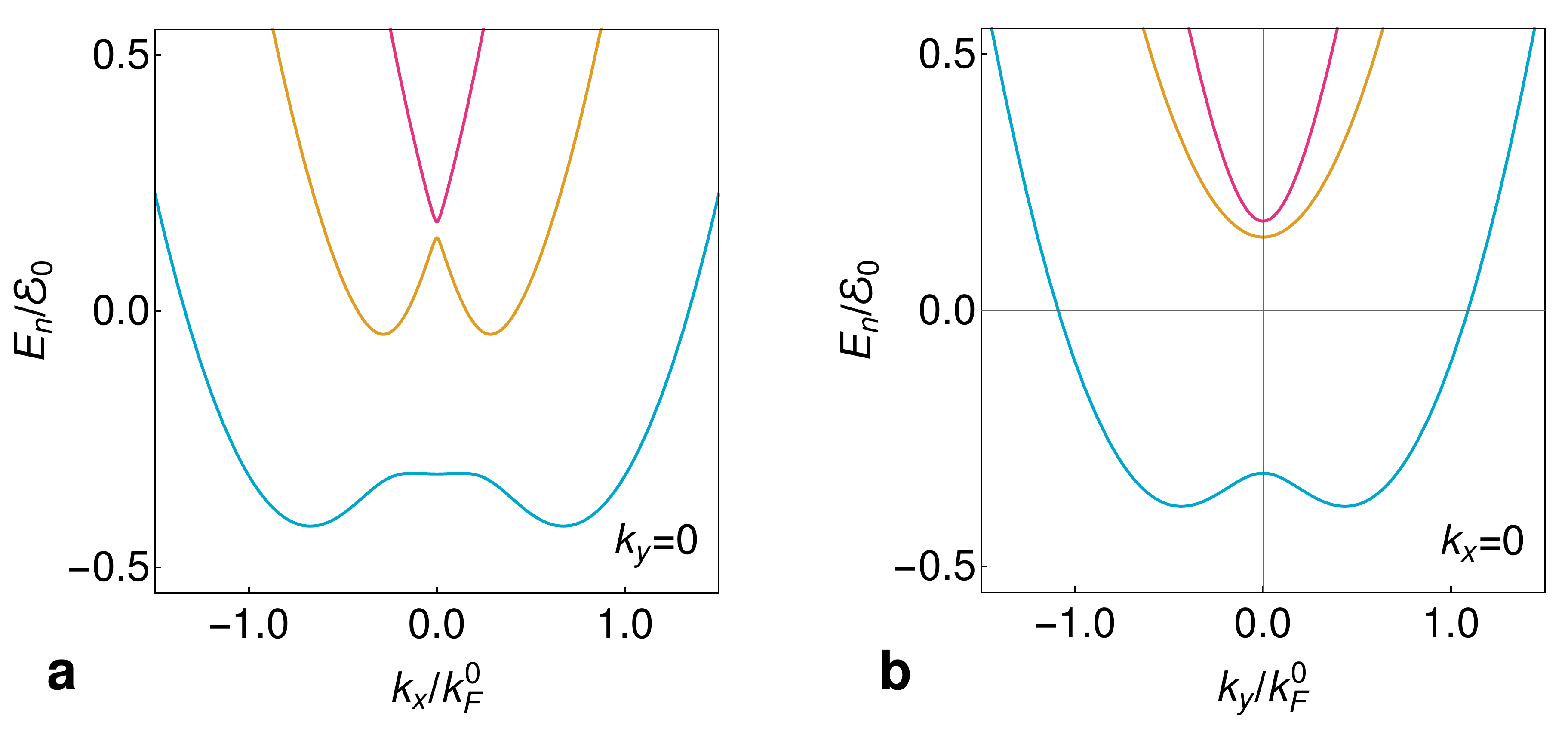}
    \caption{Energy bands of the model Hamiltonian Eq.5 of the main text including a (large) constant term $\propto \Lambda_1$. Here we have used a trigonal crystal field $\Delta=-0.2{\cal{E}}_{0}$, the additional crystal field splitting due to rotational symmetry breaking
    $\Delta_m=0.01 {\cal{E}}_{0} $, and the strengths of the orbital Rashba coupling $\alpha_{OR} = 1.0 {\cal{E}}_{0} / k_F^0$, $\alpha_m=\alpha_{OR}$. 
    For the constant term $\propto \Lambda_1$ we considered a strength $\Delta_1=0.15 {\cal{E}}_{0}$.}
    \label{figSMrev5}
\end{centering}
\end{figure*}

Even though this regime is not 
realistic,
we have explored it for completeness of our study. In Fig.\ref{figSMrev5} we display the behavior of the energy bands on the $k_y \equiv 0$ line and the  mirror symmetric line $k_x=0$. On the latter, one clearly sees that at all momenta the bands are non-degenerate, and the mirror symmetry-protected crossings discussed in the main text are absent. This is because the constant term $\propto \Lambda_1$ moves the crossings towards the center of the BZ until (for $\Delta_1=\Delta_{1,c}$ of Fig.\ref{figSMrev4}) they merge and annihilate each other.

\begin{figure*}[h!]
\begin{centering}
    \includegraphics[width=.95\textwidth]{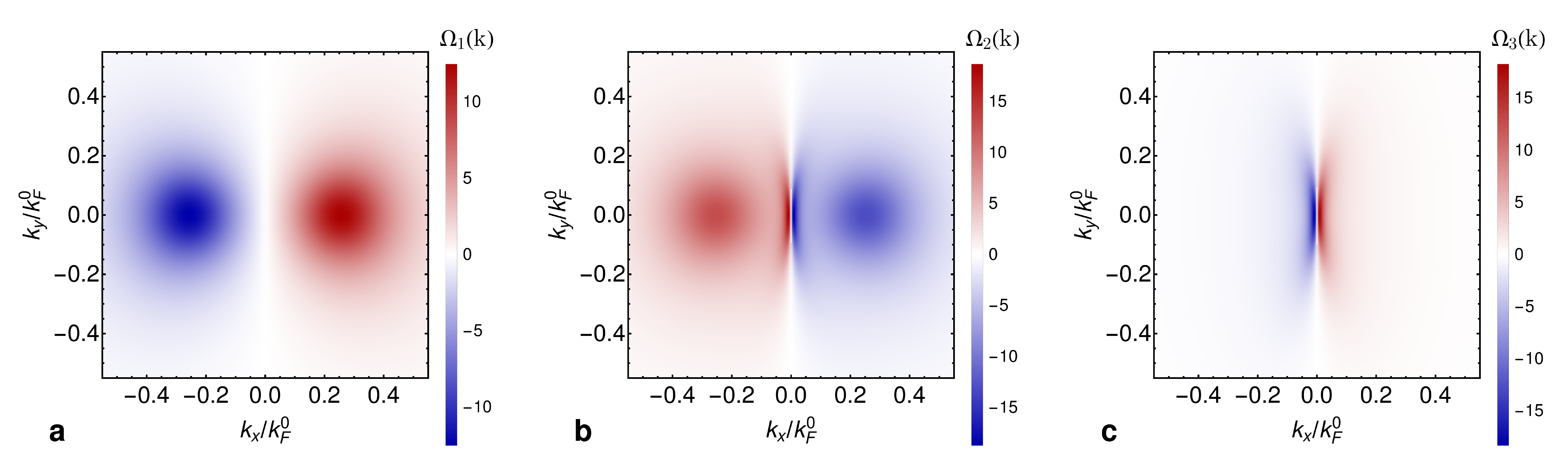}
    \caption{Density plot of the Berry curvature for the model Hamiltonian Eq.5 of the main text supplemented by a (large) constant term $\propto \Lambda_1$. The annihilation of the mirror symmetry-protected crossings results in the absence of pinch points in the Berry curvature. We have used the same parameter set of Fig.~\ref{figSMrev5}. }
    \label{figSMrev6}
\end{centering}
\end{figure*}

\begin{figure*}[h!]
\begin{centering}
    \includegraphics[width=0.35\textwidth]{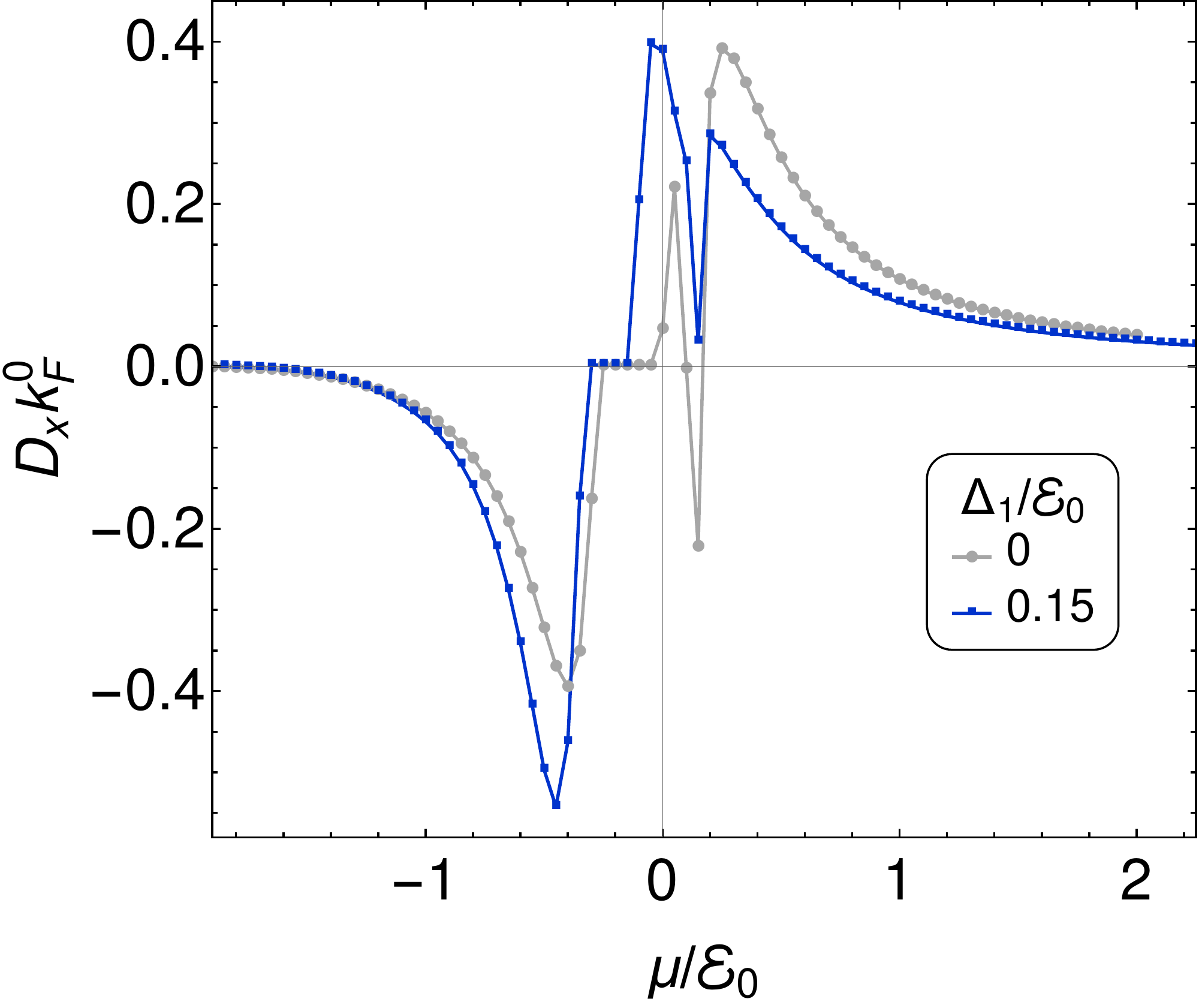}
    \caption{Behavior of the Berry curvature dipole as a function of the chemical potential for the model Eq.5 of the main text and a generalized model which includes the constant term $\propto \Lambda_1$ with a strength  $\Delta_1 \gg \Delta_m$ and that of the orbital Rashba coupling $\alpha_{R}= \alpha_m= 2.0 {\cal{E}}_{0} / k_F^0$. All other values of the model parameters are same as in Fig.~\ref{figSMrev5}.}
    \label{figSMrev7}
\end{centering}
\end{figure*}

This qualitative difference in the electronic bands is reflected in different characteristics of the Berry curvature whose behavior, in the $\Delta_1 \gg \Delta_m$ regime, is shown in Fig.~\ref{figSMrev6}. 
The Berry curvature is characterized by hot-spots with a dipolar profile, as mandated by time-reversal symmetry. However, due to the absence of mirror symmetry-protected crossings, the Berry curvature pinch points cannot exist. 
We note that the presence of hot-spots guarantees values of the Berry curvature dipole [see Fig.~\ref{figSMrev7}] of the order of inverse of the Fermi wavevector $k_F^0$, and thus of the same order of magnitude of the $\Delta_1 \equiv 0$ case. 
This proves that the size of the Berry curvature dipole is a robust feature of our symmetry-based model Hamiltonian. 

\section{Additional figures}
\begin{center}
\begin{figure*}[htb]
\includegraphics[width=0.96\textwidth]{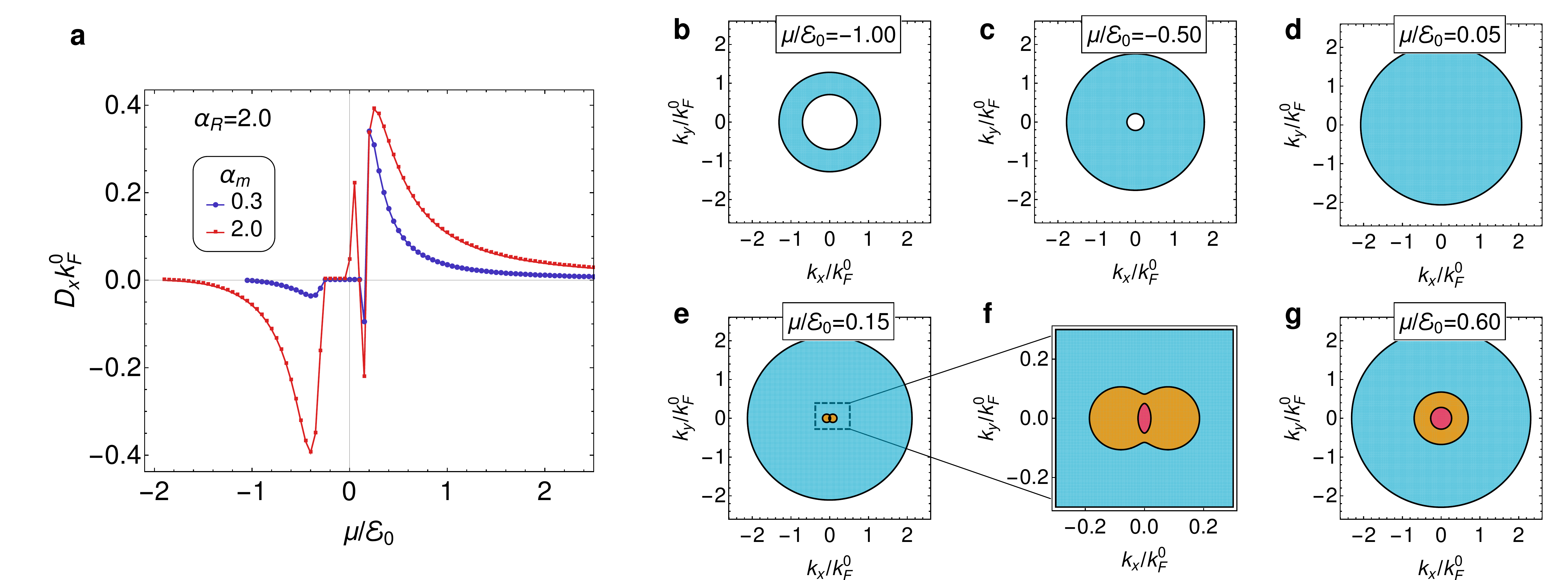} 
\protect\caption{{\bf a} Berry curvature dipole as a function of  $\mu$  for two values of $\alpha_m$ (namely $\alpha_m= 0.3 {\cal E}_0/k_F^0$  and  $\alpha_m=2.0 {\cal E}_0/k_F^0$) and for $\alpha_R=2.0{\cal E}_0/k_F^0$, $\Delta=-0.2{\cal E}_0$ and  $\Delta_m=0.01{\cal E}_0$. {\bf b},{\bf c},{\bf d},{\bf e},{\bf f} display the Fermi lines and the band-resolved occupied regions in momentum space  for $\alpha_R=2.0{\cal E}_0/k_F^0$ and  $\alpha_m=0.3 {\cal E}_0/k_F^0$. Note that the absence of the Berry curvature dipole peak at negative values of the chemical potential for $\alpha_m=0.3 {\cal E}_0/k_F^0$ is consistent with the fact that the Fermi lines retain an almost symmetric profile in this region.}
\label{figSM4}
\end{figure*}
\end{center}

\begin{center}
\begin{figure*}[htb]
\includegraphics[width=0.94\textwidth]{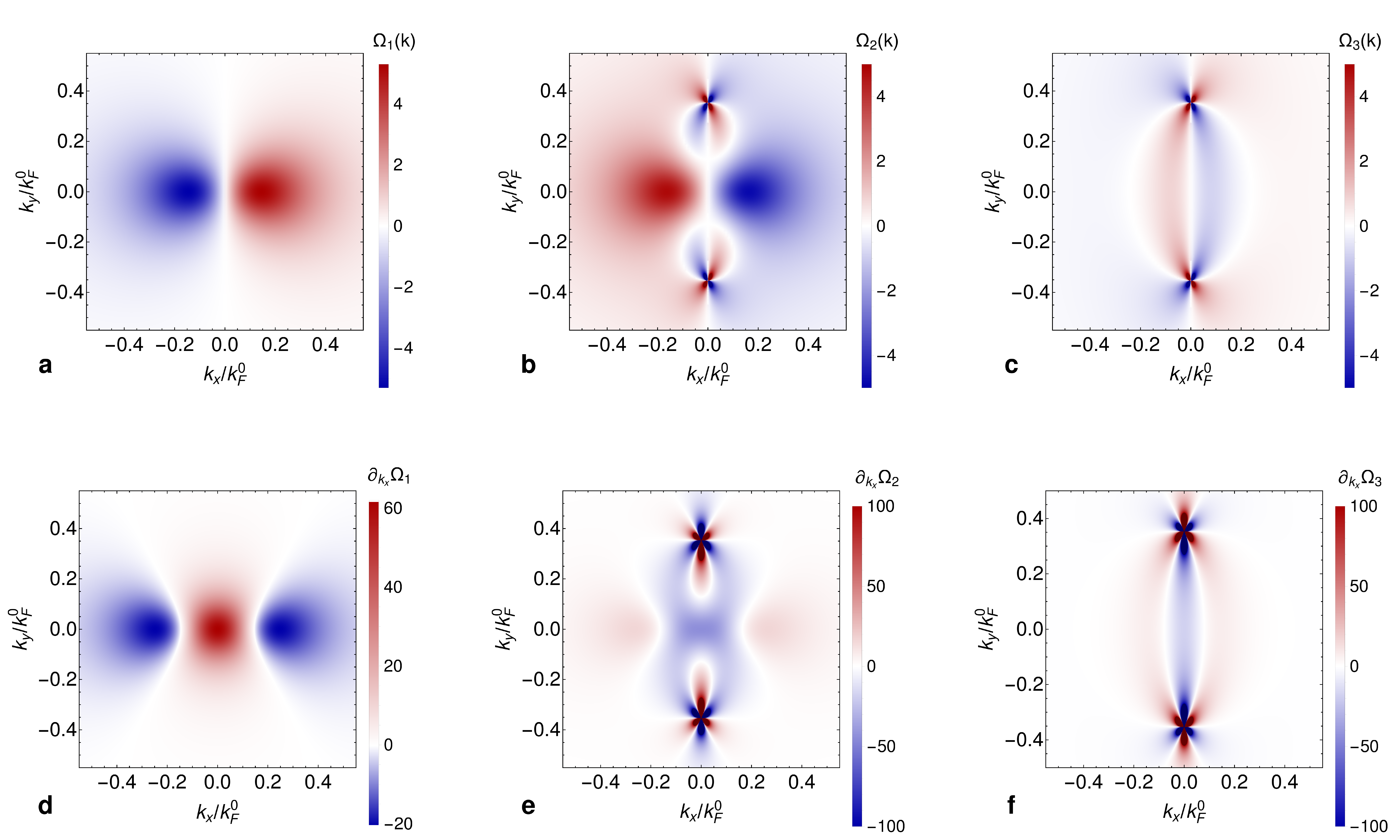} 
\protect\caption{ Density plots of Berry curvature and Berry curvature dipole for $\Delta=-0.2{\cal E}_0; \Delta_M=0.12{\cal E}_0, \alpha_R=1.0{\cal E}_0/k_F^0, \alpha_M=0.5{\cal E}_0/k_F^0$. }
\label{figSM5}
\end{figure*}
\end{center}

Fig.~\ref{figSM4} shows a comparison of the behavior of the Berry curvature dipole as a function of the chemical potential by changing the relative strength between the three-fold rotational symmetric interorbital mixing $\propto \alpha_R$ and the term $\propto \Lambda_7$ only allowed in the absence of rotational symmetries parametrized by $\alpha_m$. 
Finally Fig.~\ref{figSM5} shows the local Berry curvature and the Berry curvature dipole density for a set of parameters different from the one reported in the main text.

